\documentclass[aps,pra,reprint, amsmath, amssymb,floatfix,superscriptaddress]{revtex4-1}
 
\usepackage{bm}
\usepackage{bbm}
\usepackage[retainorgcmds]{IEEEtrantools}
\usepackage{graphicx}
\usepackage{mathrsfs}
\usepackage{amsmath}
\usepackage{amsfonts}
\usepackage{amssymb}
\usepackage{color,xcolor}
\usepackage{times,txfonts}
\usepackage{nicefrac}
\usepackage{ragged2e}
\usepackage{float}
\usepackage{braket}
\usepackage{tikz}

\usepackage[colorlinks=true,linkcolor=blue,urlcolor=blue,citecolor=blue,pdfusetitle]{hyperref}

\usepackage{blindtext}

\newcommand{\tr}{\mathrm{tr}}

\begin{document}

\title{Framework for fluctuating times and counting observables in stochastic excursions}
\date{\today}
\author{Guilherme Fiusa}
\email{gfiusa@ur.rochester.edu}
\affiliation{Department of Physics and Astronomy, University of Rochester, Rochester, New York 14627, USA}
\author{Pedro E. Harunari}
\affiliation{Complex Systems and Statistical Mechanics, Department of Physics and Materials Science, University of Luxembourg, 30 Avenue des Hauts-Fourneaux, L-4362 Esch-sur-Alzette, Luxembourg}
\affiliation{Aix Marseille Université, CNRS, CINAM, Turing Center for Living Systems, 13288 Marseille, France}
\author{Abhaya S. Hegde}
\affiliation{Department of Physics and Astronomy, University of Rochester, Rochester, New York 14627, USA}
\author{Gabriel T. Landi}
\affiliation{Department of Physics and Astronomy, University of Rochester, Rochester, New York 14627, USA}

 \begin{abstract}
Many natural systems exhibit dynamics characterized by alternating phases or recurring sets of states. 
Describing the fluctuations of such systems over stochastic trajectories is necessary across diverse fields, from biological motors to quantum thermal machines. In an accompanying Letter, we introduced the notion of stochastic excursions---a framework to analyze out of equilibrium processes via sub-trajectories. Through counting observables, this framework captures finite-time fluctuations and trajectory-level behavior, which provides insights into thermodynamical trade-offs between thermodynamic quantities of interest, such as entropy production and dynamical activity. In this work, we enhance this formalism by providing a suite of technical results on how to efficiently compute excursion-related quantities. Our analytical results provide explicit formulas for general moments of counting variables and excursion duration, as well as their covariance and conditional moments. We show that excursion counting statistics recover full counting statistics results, and uncover a relation between fluctuations of counting observables at the single-excursion level and the steady state diffusion coefficient (noise). We also discuss a fluctuation theorem for individual excursions and show that it implies a modified thermodynamic uncertainty relation at the excursion level. In addition, we explore how analyzing excursions and using the results developed here can yield insights into three problems of interest: the three-qubit absorption refrigerator, cellular sensing, and birth-and-death processes.
 \end{abstract}

\maketitle{}

\section{Introduction}
A signature of many natural phenomena is the alternating behavior between an ``inactive'' phase $A$ and an ``active'' phase $B$.
These kinds of systems transition from $A$ to $B$ when activated (e.g., by an external energy input), perform some arbitrary task in $B$, and eventually return to $A$.
This type of trajectory is characterized by the concept of \emph{stochastic excursions}, which is a framework for analyzing sub-trajectories of  {out of equilibrium Markov jump processes}~\cite{Fiusa2025}. 
 {Systems with a recurring set of states are naturally phrased in the stochastic excursion language}.
Problems of this form appear across diverse fields, including biology~\cite{bergPhysicsChemoreception1977, Mora2010, Mora2010, Harvey2023, Fancher2020, Barato2015, Lang2014, Barato2016, Barato2014}, molecular machines~\cite{Seifert2012, Kumar2010, Kolomeisky2007, Murugan2014}, 
heat engines and physics~\cite{Brandner2015, Pietzonka2018, Prech2024, Fiusa2024, Hegde2025}, chemistry~\cite{bialekPhysicalLimitsBiochemical2005}, and applied mathematics~\cite{kleinrock1974queueing, Lindley1952, gross2011fundamentals, cohen1969single, Bhat2008}.  

The concept of excursions has a long history in the theory of stochastic processes, where it is classically defined for trajectories that leave and return to a single reference state~\cite{Blumenthal1992}. Our framework rests on the a similar filtering of trajectories, but differs in two essential respects: it is formulated for Markov jump processes rather than Brownian or Langevin dynamics, and it is built around counting observables, which are naturally defined through the transitions of a discrete-state process. We also consider the system leaving/return to a set of states, rather than a single one.
Most theoretical frameworks proposed to study such systems are model-based and have focused on questions related to first-passage times~\cite{Yao1985, Rubino1989, Kook1993, Takacs1974, Garrahan2017, Bebon2024}. Those problems can be recast as questions related to excursion times.
Beyond durations, several works have also characterized time-integrated functionals, for example, the distribution of the area under Brownian excursions, avalanches, and first passage analogues~\cite{Chung1976, Majumdar2005, Majumdar2008, Majumdar2015, Takacs1991, Mazzolo2017, Zapperi2005, Papanikolaou2011, Baldassarri2021, Neri2019, Roldan2019, Meerson2020, Abundo2017, Kearney2007}.
The excursion framework we propose can also be viewed as a killed Markov process supplemented with transition-resolved counting observables and initial/final boundary weights, from which we further extract the joint statistics with the excursion duration and the connection to full counting statistics, which is related to
 reward processes with absorbing states~\cite{tan2021markovrewardsprocessesimpulse}.

This work is a longer companion to the Letter~\cite{Fiusa2025}, where we introduced the formalism, showcased its ubiquity across numerous problems of interest, and discussed the main results of this framework. 
Our goal in this accompanying paper is to construct the formalism and provide the tools and technical results to efficiently compute excursion-related quantities. 
 {At the core of our framework lies the concept of \emph{counting observables} and their statistics during excursions.
These observables are constructed by counting the number of times a specific transition took place within an excursion, and appending the appropriate weights. 
A wide range of observables such as dynamical activity and thermodynamic quantities, such as heat, work, currents, entropy production, are constructed this way.
In this longer companion paper, we provide a toolkit of simple formulas to evaluate the statistics of excursion durations, counting observables, and their interplay. 
All formulas are written as functions of elements of the stochastic transition matrix.
We also show that a fluctuation theorem specialized for excursions naturally appear from the probability distribution of counting observables and local detail balance. We use this fluctuation theorem to derive a thermodynamic uncertainty relation at the excursion level.}
We also expand on applications by showing how we use the formalism developed herein to characterize fluctuations in the three-qubit absorption refrigerator, a cellular sensing problem, and a birth-and-death process.

The organization of the present work is as follows. In Sec.~\ref{sec:formalism}, we introduce the formalism of stochastic excursions and counting observables. In Sec.~\ref{sec:main_results}, we explore the statistics of counting observables and present our main technical results, including analytical formulas for the moments of counting observables, excursion durations, and their correlations. Furthermore, we use the results derived in the section to connect our framework to full counting statistics~\cite{landi2024a, Campisi2011, Esposito2007, Esposito2009, Brandes2008, Saito2008}, providing formulas for average currents and diffusion coefficients.
In Sec.~\ref{sec:explicit-formulas}, we derive explicit results for excursion activity and specific transition counts. 
In Sec.~\ref{sec:FT}, we show a fluctuation theorem~\cite{Jarzynski1997, Crooks1999, Talkner2007a, Talkner2007b, Talkner2008, Hasegawa2019a, Hasegawa2019b, Campisi2009, Campisi2010, Campisi2014, Andrieux2009, Jarzynski2004, Jarzynski2011, Seifert2012, Esposito2009, Harris2007, Horowitz2019, Nakamura2010} for individual excursions and  {use it to derive a thermodynamic uncertainty relation~\cite{Gingrich2016, Hasegawa2019b, Potts2019} where the fluctuations of observables along excursions
in forward and backward trajectories are always tied via entropy production.} In Sec.~\ref{sec:applications}, we apply our framework to three problems of interest: the three-qubit absorption refrigerator, cellular sensing, and birth-and-death processes. Finally, in Sec.~\ref{sec:conclusion}, we discuss the broad implications of our findings and outline future directions.

\section{Formalism}
\label{sec:formalism}
Consider a system that occupies a discrete alphabet of states, evolving stochastically according to a Markovian classical master equation. The time evolution of the occupation probability $p_x$ is given by 
\begin{equation}\label{M}
    \frac{dp_x}{dt} = \sum_\ell \sum_{y\neq x} W_{\ell xy} p_y - \Gamma_x p_x,\qquad \Gamma_x = \sum_\ell \sum_{y\neq x} W_{\ell yx},
\end{equation}
where $\ell$ labels different transitions between a pair of states, each with rate $W_{\ell xy}$, and will be omitted if there is only one transition per pair. These different transitions represent mechanisms taking the system from one state to another, and can represent e.g. different reservoirs mediating the same transformation or different chemical reactions.
Eq.~\eqref{M} can be written in vector notation as 
\begin{equation}\label{M_vec}
    \frac{d|p\rangle}{dt} = \mathbb{W}|p\rangle,
\end{equation}
where $|p\rangle$ is a vector with entries $p_x$ and 
\begin{equation}\label{mathbb_W}
    \mathbb{W} = W- \Gamma = \begin{cases}
        W_{xy} & x\neq y \\
        -\Gamma_x & x = y.
    \end{cases}
\end{equation}
Here and throughout $W$ is the matrix with entries $W_{xy} = \sum_\ell W_{\ell xy}$ for $x\neq y$, and zeros in the diagonal; conversely, $\Gamma$ is the diagonal matrix with $\Gamma_x$ in the diagonals. 
Assuming irreducibility of the state space, Eq.~\eqref{M} has a unique steady state $|p^{\rm ss}\rangle$ which is the solution of $\mathbb{W}|p^{\rm ss}\rangle =0$. 
We also define a vector $\langle 1| = (1,1,1,\ldots,1)$ containing all ones. 
Normalization then implies that 
$\langle 1|\mathbb{W} = 0$; i.e.,  the columns of $\mathbb{W}$ add to zero.

Eq.~\eqref{M} describes the dynamics at the level of the ensemble. 
Conversely, we can also describe the process in a single stochastic trajectory, where the system spends a random amount of time in a given state before jumping on to the next one. 
Such stochastic trajectories are comprised of three elements: (i) a set of states $x_1\to x_2\to x_3\to \ldots$ that the system navigates; (ii) residence times $\tau_i$, representing how much time the system spent in state $x_{i}$ before jumping to $x_{i+1}$; 
and (iii) transition sources $\ell_{i}$ that generated the transition from $x_i\to_{\ell_i} x_{i+1}$, e.g. different reservoirs. 

\subsection{Regions $A$ and $B$}

\begin{figure}
    \centering
    \includegraphics[width=0.49\linewidth]{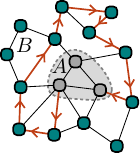}
    \caption{A system undergoes stochastic dynamics on a finite set of states. The states are split into two regions, called $A$ (grey) and $B$ (green). 
    An excursion starts with a transition from $A\to B$ and ends with the first transition back from $B\to A$ (red arrows). 
     The states of region $A$ can be spread out across the space state.}
    \label{fig:drawing}
\end{figure}
Suppose we split the alphabet of states in two arbitrary regions, denoted by $A$ and $B$, see Fig~\ref{fig:drawing}. 
The transition matrix is then split into blocks, as 
\begin{equation}\label{block_W}
    \mathbb{W} = \begin{pmatrix}
\mathbb{W}_{\!\!A} & W_{\!\!AB} \\
W_{\!\!BA} & \mathbb{W}_{\!\!B}
\end{pmatrix},
\end{equation}
where, for the diagonal blocks, 
$\mathbb{W}_{\!A} = W_A- \Gamma_A$ and $\mathbb{W}_{\!B} = W_B - \Gamma_B$. 
This decomposition is done by first organizing the matrix with states from the $A$ region, where the first $|A| \times |A|$ block defines the matrix $\mathbb{W}_{\!\!A}$.
The rest naturally follows from this structure.
Note that $\mathbb{W}_{\!A}$ and $\mathbb{W}_{\!B}$ are not actual stochastic matrices (their columns do not add up to zero) since the factors of $\Gamma_A$ and $\Gamma_B$ also contain transitions from $A\to B$ and $B\to A$. 
The steady state acquires a similar block decomposition structure
\begin{equation}\label{pss_blocks}
    |p^{\rm ss}\rangle = \begin{pmatrix}
        |p_A^{\rm ss}\rangle \\[0.2cm]
        |p_B^{\rm ss}\rangle
    \end{pmatrix},
\end{equation}
as well as the vector $\langle 1| = ( \langle 1_A| ~~~\langle 1_B|)$. 
This provides the following identities
\begin{equation}\label{identities}
\begin{aligned}
    \mathbb{W}_{\!A} |p_A^{\rm ss}\rangle &= - {W}_{AB}|p_B^{\rm ss}\rangle, 
    \qquad 
    {W}_{BA} |p_A^{\rm ss}\rangle = - \mathbb{W}_{\!B}|p_B^{\rm ss}\rangle ,
    \\[0.2cm]
    \langle 1_A|\mathbb{W}_{\! A} &= - \langle 1_B | {W}_{BA},
    \qquad 
    ~~\langle 1_A|{W}_{AB} = - \langle 1_B | \mathbb{W}_{\!B},    
\end{aligned}
\end{equation}
that are often used throughout the text.

\subsection{Excursions}

\emph{Excursions} are a filtering of stochastic trajectories where the first transition is constrained to be from $A$ to $B$ and the last transition is from $B$ back to $A$. This renders the trajectory restricted to $B$ with the added initial and final transitions that rely on $A$.
The system might spend an arbitrary amount of time in $B$ and navigate over an arbitrary number of states. A schematic illustration of an excursion is given in Fig.~\ref{fig:drawing}.

We first consider excursions that start in a specific state $|x_A\rangle$ and end in an arbitrary state $|y_A\rangle$ in region $A$. 
The probability for such an excursion reads 
\begin{equation}\label{Pexc}
    P_{\tt exc} = C_{x_A \to y_A} W_{\ell_N y_A z_N} e^{-\Gamma_{z_N}\tau_N} W_{\ell_{N-1} z_N z_{N-1}} \ldots  e^{-\Gamma_{z_1}\tau_1} W_{\ell_0 z_1 x_A},
\end{equation}
where $z_i$ are all states within $B$ and $C_{x_A \to y_A}$ is a normalization constant  {that in general depends on the states that the excursion begins and ends (see below) \footnote{It is also possible to compute Eq.~\eqref{Pexc} without conditioning on any particular initial or final state, it amounts to summing (marginalizing) over all states $x_A$ and $y_A$.}}.
The factors of $e^{-\Gamma_x \tau}$ represent the residence times in state $x$ before jumping, while $W_{\ell x'x}$ represent the transition rates between the states $x$ and $x'$ due to $\ell$. 
The number of states $z_1,z_2,\ldots,z_N$ that the system visits in $B$ is random but must be at least one, so the smallest possible excursion will have the form $x_A \to z_1 \to y_A$.

The duration of an excursion is a random variable, which we denote by $\hat{T}$. 
This is a first-passage time.
Its PDF can be obtained directly from Eq.~\eqref{Pexc} as  {
\begin{equation}
    P\big(\hat{T} = t\big) = E\Biggl[\delta\Bigg(t - \sum_{j=1}^{\hat{N}} \tau_j\Bigg)\Biggr].
\end{equation}}
The calculation is done in Appendix~\ref{app:proofs-tilted} and results in 
\begin{equation}\label{Pt}
    P\big(\hat{T} = t\big) = ~  C_{x_A\to y_A}\langle y_A|W_{AB} e^{\mathbb{W}_{\!B} t} W_{BA}|x_A\rangle.
\end{equation}
This is a minor generalization of known results on first passage times~\cite{skinner2021}. The factor of $\exp{\{\mathbb{W}_{\!B} t\}}$ takes into account the time spend in region $B$.

Now, we provide intuition behind the normalization constant $C_{x_A\to y_A}$ and show its operational meaning.
Let  {$P[x_A \to z_B | x_A]$} denote the probability that an excursion starts with a jump $x_A \to z_B$ given that the it starts in $x_A$:
\begin{equation}\label{C_PzBxA}
    P[x_A\to z_B \vert x_A] = \frac{\braket{z_B|W_{BA}|x_A}}{\braket{1_B|W_{BA}|x_A}},
\end{equation}
where $\bra{1_B}$ is a vector with all entries equal to 1.
The denominator $\mathcal{K}_{BA}^{x_A}: =\braket{1_B|W_{BA}|x_A}$ ensures that the probability is properly normalized. 
We see that it corresponds to the dynamical activity from state $x_A$ to region $B$; more specifically, it is the average number of transitions per unit time from $x_A$ to $B$, and represents how frequently excursions occur stemming from a specific state $x_A$.

Similarly, let $P[y_A \mid z_B]$ denote the conditional probability that the excursion ends in $y_A$ given that, when it began, it entered region $B$ through state $z_B$ (notice that this is now independent of $x_A$). It is given by 
\begin{equation}\label{C_PyAzB}
    P[y_A \mid z_B] = \braket{y_A|W_{AB}M|z_B},
\end{equation}
where $M:=-\mathbb{W}_B^{-1}$ is the time integral of the factor $\exp{\{\mathbb{W}_{\!B} t\}}$, which accounts for the variable time that the system spends in region $B$.
Multiplying Eqs.~\eqref{C_PzBxA} and~\eqref{C_PyAzB} and summing over all states $\ket{z_B}$ yields 
the probability that an excursion ends in $y_A$ given that it started in $x_A$,
\begin{equation}
\label{p-exc-xa-ya}
    P[x_A\to y_A \vert x_A] = \frac{\braket{y_A|W_{AB} M W_{BA}|x_A}}{\mathcal{K}_{BA}^{x_A}}.
\end{equation}
The quantity in the numerator is precisely $C_{x_A\to y_A}^{-1}$. 
Thus we arrive at
\begin{equation}\label{C_interpretation}
    C_{x_A\to y_A}^{-1} = P[x_A \to y_A | x_A] \mathcal{K}_{BA}^{x_A}.
\end{equation}
We therefore see that $C_{x_A\to y_A}^{-1}$ is related to the probability of the excursion ending in $y_A$ conditioned on its start in $x_A$, and the dynamical term $\mathcal{K}_{BA}^{x_A}$. 
This attributes physical meaning to the normalization constant $C_{x_A \to y_A}$  {that was first seen in Eq.~\eqref{Pexc}} and appears in several calculations throughout this text. Note that $C_{x_A\to y_A}$ has units of time, and it represents a relevant timescale of excursions.

\subsection{Counting observables}
 {The main feature of the framework proposed here is the characterization of statistics of counting observables within excursions. We first provide the definition of those observables and then discuss what happens once we restrict those to excursions, providing a few examples of important quantities.} 

 {For a given Markov jump process, we define a random variable $\hat{N}_{\ell xy}$ that counts the number of times the transition $y \to_\ell x$ was observed in one excursion.}
A \emph{counting observable} is then defined as a linear combination of these random variables:
\begin{equation}\label{counting_observable}
    \hat{Q} = \sum_\ell \sum_{x,y} \nu_{\ell xy} \hat{N}_{\ell xy},
\end{equation}
with generic weights $\nu_{\ell xy}$~\cite{landi2024a}.
 {The counting observable registers the first transition of the excursion $A \to B$, all transitions within $B$, and the very last transition $B \to A$.}
Counting observables are defined irrespective of which state the excursion begins and ends. 

With different choices of weights $\nu_{\ell xy}$ we can assess a wide variety of physical quantities. 
A typical quantity of interest is the number of times a specific transition $z'\to z$ inside region $B$ occurs. This is obtained by setting $\nu_{\ell zz'}=1$ and all other weights as zero. We provide analytical formulas for this case in Sec.~\ref{subsec:specific}.
Similarly, $\nu_{\ell xy}=1 \, \forall x,y$ for some $\ell$, can address the number of transitions of a given type. This can evaluate, for example, the activity of some hot reservoir in contrast to some other cold reservoir.

Another simple example of a counting observable is counting how many transitions occur in total. 
This is achieved by setting $\nu_{\ell xy} = 1$ in Eq.~\eqref{counting_observable}.
This counting observable is referred as the \emph{excursion activity} $\hat{\mathcal{A}}$, it is related to the more usual dynamical activity (freneticity) which characterizes the number of transitions per unit time. 
The difference is that instead of characterizing activity per unit time, excursion activity accounts for activity per excursion. We provide analytical formulas for the excursion activity in Sec.~\ref{subsec:activity}. 
 {The dynamical activity has attracted a lot of interest because it plays a key role in uncertainty relations~\cite{Di_Terlizzi2018, Nishiyama2024, Vo2022}. The simple relation between dynamical and excursion activity could be an important tool to explore excursions in the context of uncertainty relations.}

Linear counting observables are also typically used to describe thermodynamic currents. 
In this case, the corresponding weights must be anti-symmetric $\nu_{\ell xy} = -\nu_{\ell yx}$~\cite{Wachtel2015}.
If transition rates are generated by thermal reservoirs, local detailed balance is satisfied:
\begin{equation}\label{detailed_balance}
    \frac{W_{\ell xy}}{W_{\ell yx}} = e^{-\beta_\ell (E_x - E_y)},
\end{equation}
where $E_{x/y}$ are the energies of levels $x/y$ and $\beta_\ell$ is the inverse temperature of the reservoir associated with transition $\ell$. 
The counting observable representing the entropy produced in the process,  {given by the log of Eq.~\eqref{detailed_balance}}, can then be built with weights $\nu_{\ell xy} = -\beta_\ell (E_x-E_y)$, which results in 
\begin{equation}\label{entropy_production_counting_observable}
    \hat{\Sigma} := -\sum_\ell \sum_{x,y} \beta_\ell(E_x-E_y) \hat{N}_{\ell xy}.
\end{equation}
Other thermodynamic quantities can be constructed similarly. In Sec.~\ref{sec:FT}, we show that the entropy production satisfies a fluctuation theorem at the excursion level, where negative entropy production is exponentially suppressed, and use this result to derive a modified version of the thermodynamic uncertainty relation of Ref.~\cite{Hasegawa2019b}.

\section{Statistics of counting observables within an excursion}\label{sec:main_results}

Having set the formalism in place, we now characterize the statistics of counting observables.
Consider a set of $r$ counting observables $\hat{Q}_\alpha$, defined as in Eq.~\eqref{counting_observable}, 
each with its own set of weights $\nu_{\ell xy}^{\alpha}$. 
Introduce tilted matrices 
\begin{equation}\label{tilted_transition_matrix}
    (\mathbb{W}_{\bm{\xi}})_{xy} =
    \begin{cases}
    \sum_\ell W_{\ell xy} e^{-i \sum_\alpha \nu_{\ell xy}^\alpha \xi_\alpha} & x\neq y
    \\[0.2cm]
    -\Gamma_x & x = y,
    \end{cases}
\end{equation}
with $r$ counting fields $\xi_\alpha$.
For an excursion starting in $x_A$ and ending in $y_A$, the joint probability that each observable $\hat{Q}_\alpha$ takes on a value $q_\alpha$ \emph{and} that total excursion time is $\hat{T} = t$ reads 
\begin{equation}\label{Pqt}
    P(\bm{q},t) = C_{x_A \to y_A} \int\limits_{-\infty}^\infty \frac{d\xi_1\ldots d\xi_r}{(2\pi)^r} \langle y_A| W_{AB\bm{\xi}} e^{\mathbb{W}_{\!B\bm{\xi}}t} W_{BA\bm{\xi}} |x_A\rangle e^{i \bm{q}\cdot \bm{\xi}},
\end{equation}
where $\bm{q} = (q_1, \ldots, q_r)$.
All other calculations follow from this result.
The proof of this result is in Appendix~\ref{app:proofs-joint-prob}, and the intuition is as follows. 
The counting fields $\bm{\xi} = (\xi_1,\ldots,\xi_r)$ pick up all events that occur during an excursion. 
We first count the starting jump from $A\to B$ using $W_{BA\bm{\xi}}$. 
Then we count an arbitrary number of jumps within $B$ using $\exp \{ \mathbb{W}_{\!B\bm{\xi}}t \}$. And finally we count the jump from $B\to A$ with $W_{AB\bm{\xi}}$. 
With the tilted transition matrix~\eqref{tilted_transition_matrix}, each transition $y\to_\ell x$ picks up the correct factor of $\nu_{\ell xy}^{\alpha}$ for each counting field. The integral over all $\xi_\alpha$ in Eq.~\eqref{Pqt} then takes us from the characteristic function to the actual probability distribution. 

Marginalizing Eq.~\eqref{Pqt} over all counting fields yields us back the excursion time distribution~\eqref{Pt}.
Conversely, marginalizing over the excursion duration yields us the joint distribution of the counting observables within the excursion: 
\begin{equation}\label{Pq_xy}
    P(\bm{q}) = C_{x_A\to y_A} \int\limits_{-\infty}^\infty \frac{d\xi_1\ldots d\xi_r}{(2\pi)^r} \langle y_A| W_{AB\bm{\xi}}M_{\bm{\xi}} W_{BA\bm{\xi}} |x_A\rangle e^{i \bm{q}\cdot \bm{\xi}},
\end{equation}
where we defined $M_{\bm{\xi}} := -\mathbb{W}_{\!B\bm{\xi}}^{-1}$.

The results stated in this section so far rely on a specific choice of initial and final states, $x_A$ and $y_A$, within region $A$. Very often, one is interested only in the statistics of excursions, irrespective of where exactly it starts or ends.
This is true, for example, when we want to analyze the long-time statistics over many excursions. 
In this case, results from Eqs.~\eqref{Pt},~\eqref{Pqt}, and~\eqref{Pq_xy} remain valid provided that initial states $\ket{x_A}$ are sampled from the the steady state distribution  $|p_A^{\rm ss}\rangle$ [Eq.~\eqref{pss_blocks}], and we sum over all possibilities for the final state, which amounts to replacing $\langle y_A|$ by $\langle 1_A|$, which is the vector of length $|A|$ with all entries equal to 1. 

\subsection{Characteristic function}

Given the formulas to characterize the statistics of counting observables within an excursion, we now go a step further and analyze the moments of generic counting observables. 
Since the integrand in Eq.~(\ref{Pq_xy}) is the characteristic function
\begin{equation}\label{CF}
    G(\bm{\xi}) = C_{x_A \to y_A} \langle y_A| W_{AB\bm{\xi}} M_{\bm{\xi}} W_{BA\bm{\xi}} |x_A\rangle,
\end{equation}then the joint characteristic function of all counting observables and the excursion time is given by
\begin{equation}
\begin{aligned}
    G(\bm{\xi},\omega) &= \int dq_1\ldots dq_r \int\limits_0^\infty dt ~e^{-i (\bm{q}\cdot\bm{\xi} + \omega t)} P(\bm{q},t) 
    \\[0.1cm]
    &= C_{x_A \to y_A} \langle y_A| W_{AB\bm{\xi}} (-\mathbb{W}_{\!B\bm{\xi}}-i\omega)^{-1} W_{BA\bm{\xi}} |x_A\rangle.
\end{aligned}
\end{equation}
Series expanding in $\omega$ and $\bm{\xi}$ yields the first few moments:
\begin{equation}\label{CF_series}
\begin{aligned}
    G(\bm{\xi},\omega) =& 1 +i \sum_{\alpha=1}^r \xi_\alpha E(\hat{Q}_\alpha) - \sum_{\alpha, \beta=1}^r \frac{\xi_\alpha \xi_\beta}{2}E(\hat{Q}_\alpha \hat{Q}_\beta)
    \\[0.2cm]
    &+i\omega E(\hat{T})- \frac{\omega^2}{2}E(\hat{T}^2) - \sum_{\alpha=1}^r \omega \xi_\alpha E(\hat{T}\hat{Q}_\alpha) + \ldots. 
\end{aligned}    
\end{equation}
We first use the geometric series expansion for matrices to expand  
$(-\mathbb{W}_{\!B\bm{\xi}}-i\omega)^{-1} = M_{\bm{\xi}}+i\omega M_{\bm{\xi}}^2 -\omega^2 M_{\bm{\xi}}^3+\ldots$, which leads to 
\begin{equation}\label{CF_series_2}
\begin{aligned}
   &G(\bm{\xi},\omega) =\ C_{x_A \to y_A}\langle y_A| W_{AB\bm{\xi}} M_{\bm{\xi}} W_{BA\bm{\xi}} |x_A\rangle + i\omega  C_{x_A \to y_A}
    \\[0.2cm]&\times\langle y_A| W_{AB\bm{\xi}} M_{\bm{\xi}}^2 W_{BA\bm{\xi}} |x_A\rangle 
    -\omega^2 C_{x_A \to y_A}\langle y_A| W_{AB\bm{\xi}} M_{\bm{\xi}}^3 W_{BA\bm{\xi}} |x_A\rangle +\ldots.
\end{aligned}
\end{equation}
This result is very general. In the following, we compute the moments of quantities of interest for excursions.

\subsection{Moments of the excursion time}

The first two moments of the duration of excursions $\hat{T}$ are obtained by setting $\bm{\xi}=0$ in Eq.~\eqref{CF_series_2} and comparing with Eq.~\eqref{CF_series}:
\begin{align}\label{average_excursion_time}
    E(\hat{T}) &= C_{x_A \to y_A}\langle y_A| W_{AB} M^2 W_{BA} |x_A\rangle,
    \\[0.2cm]
    E(\hat{T}^2) &= 2C_{x_A \to y_A}\langle y_A| W_{AB} M^3 W_{BA} |x_A\rangle.   
\end{align}
And the variance of the excursion time is 
\begin{equation}\label{variance_excursion_time}
\begin{aligned}
    {\rm var}(\hat{T}) &= 2C_{x_A \to y_A} \langle y_A| W_{AB}M^3 W_{BA}|x_A\rangle - E(\hat{T})^2.
\end{aligned}    
\end{equation}
Alternatively, any moment of the excursion time can also be easily computed directly from Eq.~\eqref{Pt} and the identity 
$\int_0^\infty dt~t^n e^{\mathbb{W}_{\!B}t} = (-1)^{n+1} n! \mathbb{W}_{\!B}^{-(n+1)}$, leading to 
\begin{equation}\label{moments_excursion_time}
\begin{aligned}
    E(\hat{T}^n) &= n! C_{x_A \to y_A}\langle y_A| W_{AB} M^{n+1} W_{BA} |x_A\rangle   
\end{aligned}
\end{equation}
Interestingly, the problem of computing moments of the excursion duration reduces to simply computing powers of the matrix $M$.

\subsection{Moments of the counting observables}

For general counting observables, we first expand the tilted matrices [c.f.~\eqref{tilted_transition_matrix}] over $\bm{\xi}$:
\begin{equation}\label{tilde_expansion}
    \mathbb{W}_{\!\bm{\xi}} \simeq \mathbb{W} + i \sum_{\alpha=1}^r \xi_\alpha W^{(\alpha)} - \frac{1}{2}\sum_{\alpha,\beta}^r \xi_\alpha \xi_\beta W^{(\alpha,\beta)},
\end{equation}
where 
\begin{align}
    W_{xy}^{(\alpha)} &= \sum_\ell W_{xy}^\ell \nu_{\ell xy}^\alpha,
    \\[0.2cm]
    W_{xy}^{(\alpha,\beta)} &= \sum_\ell \nu_{\ell xy}^\alpha \nu_{\ell xy}^\beta W_{xy}^\ell.
\end{align}
Notice these are element-wise (Hadamard) products, not matrix products.
We keep the expansion up to order two since we are only interested in the average and variance. To compute higher order momenta, such as skewness or kurtosis, all one has to do is to keep higher order terms.

Moreover, this is defined for the transition matrices $W$ which have zeros in the diagonal, and not the full matrix $\mathbb{W}$ in Eq.~\eqref{mathbb_W}.
Carrying out a similar expansion for $M_{\bm{\xi}}$:
\begin{equation}
\begin{aligned}
\label{M-expansion}
    M_{\bm{\xi}} \simeq &\ M + i \sum_{\alpha=1}^r \xi_\alpha M W_B^{(\alpha)} M 
    - \frac{1}{2}\sum_{\alpha,\beta=1}^r \xi_\alpha\xi_\beta \bigg( M W_B^{(\alpha,\beta)} M 
    \\[0.2cm]
    & + M W_B^{(\alpha)} M W_B^{(\beta)} M+M W_B^{(\beta)} M W_B^{(\alpha)} M \bigg) + \ldots .
\end{aligned}
\end{equation}
Plugging this expansion in Eq.~\eqref{CF_series_2} will give us the first and second moments of $\hat{Q}_\alpha$:

\begin{widetext}
\begin{align}
\label{mean_Q}
    E(\hat{Q}_\alpha) &= C_{x_A \to y_A}~ \bra{y_A} \Big(W_{AB}^{(\alpha)} M W_{BA}^{} + W_{AB}^{} M W_{BA}^{(\alpha)} + W_{AB}^{} M W_{B}^{(\alpha)} M W_{BA}^{}\Big) \ket{x_A},
    \\[0.2cm]
    \label{second_moment_Q}
    E(\hat{Q}_\alpha \hat{Q}_\beta) &= C_{x_A \to y_A}~\langle y_A|\Bigg\{ 
    W_{AB}^{(\alpha,\beta)} M W_{BA}^{} + W_{AB}^{} M W_{BA}^{(\alpha,\beta)} + W_{AB}^{} M W_{B}^{(\alpha,\beta)} M W_{BA}^{}
    \\[0.2cm]\nonumber
    &\quad + 
    W_{AB}^{} M W_B^{(\alpha)} M W_B^{(\beta)} M W_{BA}^{}+ 
    W_{AB}^{} M W_B^{(\beta)} M W_B^{(\alpha)} M W_{BA}^{}
    +W_{AB}^{(\alpha)} M W_{BA}^{(\beta)}
    +W_{AB}^{(\beta)} M W_{BA}^{(\alpha)}
    \\[0.2cm]\nonumber
    &\quad + 
    W_{AB}^{(\alpha)}M W_B^{(\beta)} M W_{BA}^{} + 
    W_{AB}^{(\beta)}M W_B^{(\alpha)} M W_{BA}^{} + 
    W_{AB}^{}M W_B^{(\alpha)} M W_{BA}^{(\beta)} + 
    W_{AB}^{}M W_B^{(\beta)} M W_{BA}^{(\alpha)}      
    \Bigg\} |x_A\rangle. 
\end{align}
\end{widetext}
The interpretation of Eq.~\eqref{mean_Q} is intuitive: the first and second terms are the average of $\hat{Q}_\alpha$ in the transitions from $A\to B$ and then from $B\to A$. The last term is then the average of $\hat{Q}_\alpha$ while inside $B$. 
The interpretation of Eq.~\eqref{second_moment_Q} is similar albeit more technical.
 {Note that the results~\eqref{mean_Q} and~\eqref{second_moment_Q} are exact. The expansion over the counting fields is kept up to second order in Eq.~\eqref{M-expansion} because higher order terms do not contribute to the first and second momenta.
The variance is readily evaluated by
\begin{equation}\label{variance}
    {\rm var}(\hat{Q}_\alpha) = E(\hat{Q}_\alpha^2)-E(\hat{Q}_\alpha)^2,
\end{equation}where $E(\hat{Q}_\alpha^2)$ is given by Eq.~\eqref{second_moment_Q} with $\beta = \alpha$.}
From Eqs.~\eqref{mean_Q} and~\eqref{second_moment_Q} one can also readily build the covariance matrix 
\begin{equation}\label{covariance}
    {\rm cov}(\hat{Q}_\alpha,\hat{Q}_\beta) = E\big(\hat{Q}_\alpha\hat{Q}_\beta\big) - E(\hat{Q}_\alpha) E(\hat{Q}_\beta),
\end{equation}
between two counting observables. 

\subsection{Correlation between excursion duration and counting observables}

Very often the counting observables and the excursion duration are correlated. This is clearly seen in the expansion on Eqs.~\eqref{CF_series} and~\eqref{CF_series_2}.  To calculate $E(\hat{T} \hat{Q}_\alpha)$, we perform another series expansion $i\omega C_{x_A \to y_A}\langle y_A| W_{AB\bm{\xi}} M_{\bm{\xi}}^2 W_{BA\bm{\xi}} |x_A\rangle$ to first order in $\bm{\xi}$. This results in 
\begin{equation}\label{EQT}
\begin{aligned}
    E(\hat{T}\hat{Q}_\alpha) =&\ C_{x_A \to y_A}~\langle y_A| \Big\{ W_{AB}^{(\alpha)} M^2 W_{BA}^{} + W_{AB}^{} M^2 W_{BA}^{(\alpha)} 
    \\[0.2cm]
    & + W_{AB}^{} M^2 W_{B}^{(\alpha)}M W_{BA}^{} + W_{AB}^{} M W_{B}^{(\alpha)} M^2 W_{BA}^{}\Big\}|x_A\rangle.
\end{aligned}
\end{equation}
From this result and the averages of excursion duration [Eq.~\eqref{average_excursion_time}] and generic counting observables [Eq.~\eqref{mean_Q}], one can quantitatively explore the covariance between a generic counting observable and the excursion duration.
 {Moreover, results in this section provide a systematic way to evaluate the average and fluctuations of any counting observable in an excursion as well as excursion durations with simple closed-formula expressions that depend only on elements of the stochastic transition matrix~\eqref{mathbb_W}.}

\subsection{Conditional moments}

At last, another important calculation that arises in the context of excursions is when a counting observable is conditioned. 
For simplicity, let us consider a single counting observable $\hat{Q}_\alpha$. 
We first calculate $E(\hat{Q}_\alpha|\hat{T}=t)$. 
Taking the Fourier transform of Eq.~\eqref{Pqt} with respect only to $\xi$ yields the conditional characteristic function 
\begin{equation}
    \frac{\langle y_A| W_{AB\xi} e^{\mathbb{W}_{B\xi}t}W_{BA\xi} |x_A\rangle}{\langle y_A| W_{AB} e^{\mathbb{W}_{B}t}W_{BA} |x_A\rangle}.
\end{equation}
The first order term in a series expansion in $\xi$ will give us $i\xi E(\hat{Q}_\alpha|\hat{T}=t)$.
For $W_{AB\xi}$ and $W_{BA\xi}$ we use Eq.~\eqref{tilde_expansion}. 
For the exponential we use the Baker-Campbell-Hausdorff formula 
\begin{equation}
    e^{\mathbb{W}_{B\xi}t} \simeq e^{\mathbb{W}_B t} + i \xi \int\limits_0^t dt'~e^{\mathbb{W}_B(t-t')}W_B^{(\alpha)} e^{\mathbb{W}_Bt'}.
\end{equation}
Putting all the above equations together, we find 
\begin{equation}
\begin{aligned}
    E(\hat{Q}_\alpha|\hat{T}=t) =&\ \frac{C_{x_A \to y_A}}{P(t)}\Bigg\{ \langle y_A| W_{AB}^{(\alpha)} e^{\mathbb{W}_Bt} W_{BA}^{}|x_A\rangle\\[0.2cm]
    &+ \langle y_A| W_{AB}^{}e^{\mathbb{W}_Bt} W_{BA}^{(\alpha)}|x_A\rangle
    \\[0.2cm]
    & + \int\limits_0^t dt'~\langle y_A|W_{AB}^{}e^{\mathbb{W}_B(t-t')}W_B^{(\alpha)} e^{\mathbb{W}_Bt'}W_{BA}^{}|x_A\rangle\Bigg\}.
\end{aligned}
\end{equation}
This quantity is useful for inferring the value of the counting observable from the excursion time. 
The reason is because the conditional mean is exactly the optimal mean-squared predictor; i.e., the function of $\theta(\hat{T})$ that minimizes the mean-squared error between $E\Big[\big(\hat{Q}_\alpha-\theta(\hat{T})\big)^2\Big]$.

Of course, as can be seen, this quantity is more difficult to evaluate since it requires computing the exponential of $\mathbb{W}_B$ and integrating. Instead, a simpler quantity for predicting $\hat{Q}_\alpha$ is the optimal \emph{linear} predictor, given by 
\begin{equation}
\label{eq:inference}
   \theta_{\rm lin}(\hat{T}) = E(\hat{Q}_\alpha)  + \frac{{\rm cov}(\hat{Q}_\alpha, \hat{T})}{{\rm var}(\hat{T})}\Big(\hat{T} - E(\hat{T})\Big).
\end{equation}
This quantity can be straightforwardly calculated from Eqs.~\eqref{average_excursion_time},~\eqref{variance_excursion_time},~\eqref{mean_Q} and \eqref{EQT}. 
This result is a type of inference that can be obtained from the excursion framework by assessing the statistics of excursion times.
Inference of counting observables such as entropy production in hidden/coarse grained settings is a quite active area~\cite{rahavFluctuationRelationsCoarsegraining2007, boEntropyProductionStochastic2014, biskerHierarchicalBoundsEntropy2017, yuDissipationLimitedResolutions2024a, harunariWhatLearnFew2022, espositoStochasticThermodynamicsCoarse2012, ehrichTightestBoundHidden2021, blomMilestoningEstimatorsDissipation2024, liangThermodynamicBoundsSymmetry2024, Tan2021}, it would be interesting to further explore what insights Eq.~\eqref{eq:inference} can bring in this context, especially because inference of counting observables beyond entropy production is a very difficult task ~\cite{van_der_Meer2022, van_der_Meer2023}.

\subsection{Connection with full counting statistics steady state}
\label{subsec:fcs}

We now discuss how to use the results developed in this section to connect our framework with full counting statistics (FCS),  {which is a framework developed to address current fluctuations and statistics of integrated charges~\cite{schaller2014open, landi2024a, Campisi2011, Esposito2007, Esposito2009, Brandes2008, Saito2008}.} 
Let $\hat{\mathcal{Q}}$ denote a FCS counting observable. This counting observable is defined in a similar way to Eq.~\eqref{counting_observable}, but considered over a fixed time window $[0,t]$.  {That is,
\begin{equation}
    \hat{\mathcal{Q}}(t) = \sum_{\ell xy} \nu_{\ell xy} \hat{N}_{\ell xy}(t),
\end{equation}where the random variable $\hat{N}_{\ell xy}(t)$ counts how many times the transition $y \to_\ell x$ took place within the time window.}
In the limit of long time $t$, the average current and the diffusion coefficient (also called noise) of this observable are defined as
\begin{equation}\label{FCS_J_D}
    J = \lim\limits_{t\to\infty} \frac{E(\hat{\mathcal{Q}})}{t},
    \qquad 
    D = \lim\limits_{t\to\infty} \frac{{\rm var}(\hat{\mathcal{Q}})}{t}.
\end{equation} We will show next how to recover the same results but using counting observables defined per excursion.

Let us assume region $A$ is spanned by a single state, this means that excursions become statistically independent of each other. The connection between FCS results and counting observables at the excursion level comes from a concatenation of several excursions, all starting in the same state $x_A$. Let $\hat{Q}_n,~\hat{T}_n$ denote a counting observable and the excursion time during the $n$-th excursion. 
The total \emph{cycle} time of an excursion also includes the residence time $\hat{\tau}_{x,n}$ (i.e. the time the system spends in $x$ after the $(n-1)$-th excursion is over but before the $n$-th excursion has started), so it is defined as $\hat{T}_n^{\rm cyc} = \hat{T}_n + \hat{\tau}_{x,n}$.
Note that $\hat{\tau}_{x,n}$ and $\hat{T}_n$ are statistically independent. 
Finally, let $\hat{\mathcal{N}}(t)$ denote the number of excursions that took place in the interval $[0,t]$. This random variable forms a renewal process~\cite{cox1967renewal}. For sufficiently large $t$, we can then write 
\begin{equation}
\label{eq:fcs-excursion-q}
    \hat{\mathcal{Q}} \simeq \sum_{n=1}^{\hat{\mathcal{N}}(t)} \hat{Q}_{n},
\end{equation}
and the total time can be broken into a sum of cycle times:
\begin{equation}
    \sum_{n=1}^{\hat{\mathcal{N}}(t)} \hat{T}_n^{\rm cyc} \simeq  t.
\end{equation}
The error in the approximation is manifested only in the boundary terms, since at time $t$ the ($N+1$)-th excursion may be still in its course. This means it is sub-extensive and can be discarded in the long $t$ limit.

The average current and noise from Eq.~\eqref{FCS_J_D} can be obtained as
\begin{align}
    \label{fcs-current}
    J &= \frac{E(\hat{Q})}{\mu},\\
    \label{fcs-diffusion}
    D &= \frac{{\rm var}(\hat{Q})}{\mu} + \frac{\Delta^2}{\mu^3}E(\hat{Q})^2 - \frac{2E(\hat{Q})}{\mu^2}{\rm cov}(\hat{Q},\hat{T}),
\end{align}
where we defined $\mu = E(\hat{T}_n^{\rm cyc}) = E(\hat{T}) + \Gamma_x^{-1}$ and $  \Delta^2 = {\rm var}(\hat{T}_n^{\rm cyc}) = {\rm var}(\hat{T}) +\Gamma_x^{-2}$ as the mean and variance of $\hat{T}^{\rm cyc}$. Residence times $\hat{\tau}_x$ are taken to be exponentially distributed with parameter $\Gamma_x$, and since they are statistically independent from $\hat{Q}$ and $\hat{T}$, it follows that 
${\rm cov}(\hat{Q},\hat{T}^{\rm cyc})={\rm cov}(\hat{Q},\hat{T})$. We show the proof of this result in Appendix~\ref{app:proofs-J-D}.

The results in Eqs.~\eqref{fcs-current} and~\eqref{fcs-diffusion} are one of the main achievements of the excursion framework.
 {First, it provides a way to compute steady state quantities in FCS by looking at statistics of individual excursions, which are subtrajectories and therefore easier to treat.
Second, it unravels a decomposition of the FCS noise into three distinct components, shedding light into the mechanisms behind noise generation in classical stochastic dynamics.
Investigating this further, and especially under the lenses of uncertainty relations, can provide fundamental insights into the nature of current precision.}
With the tools developed in this section, such as Eqs.~\eqref{average_excursion_time},~\eqref{variance_excursion_time},~\eqref{mean_Q},~\eqref{second_moment_Q},~\eqref{variance}, and~\eqref{EQT}, the average current and the noise can be readily computed.

\section{Explicit formulas for the counting observable distribution}
\label{sec:explicit-formulas}
Having developed the stochastic excursions formalism and the technical tools in the previous sections, here we explicitly compute analytical results for the dynamical activity and for counting a set of specific transitions.

\subsection{Dynamical activity}
\label{subsec:activity}

The dynamical activity $\hat{\mathcal{A}}$ per excursion is obtained by setting weights $\nu_{\ell xy} = 1$ for all transitions. Explicitly,
\begin{equation}
\label{eq:a-activity}
    \hat{\mathcal{A}} = \sum_{\ell xy} \hat{N}_{\ell xy}.
\end{equation}
As a consequence, the tilted matrix~\eqref{tilted_transition_matrix} becomes $\mathbb{W}_{\xi} = e^{i \xi} W - \Gamma$. 
Hence $\mathbb{W}_{\!B\xi} = e^{i\xi} W_B - \Gamma_B$, while for the block off-diagonals we simply have $W_{AB\xi} = e^{i \xi} W_{AB}$ and $W_{BA\xi} = e^{i \xi} W_{BA}$.
Eq.~\eqref{CF} then becomes 
\begin{equation}\label{eq:gen-function-activity}
    G(\xi) = C_{x_A \to y_A}
    \langle y_A| W_{AB} M_{\xi} W_{BA} |x_A\rangle  e^{2 i \xi}.
\end{equation}
To evaluate $M_\xi$ we use the geometric series expansion for matrices, which yields
\begin{equation}
\begin{aligned}
  M_\xi &=  -\mathbb{W}_{\!B\xi}^{-1}  = (\Gamma_B - e^{i\xi} W_B)^{-1}
    \\[0.2cm]
    &= \Gamma_B^{-1} + e^{i\xi}\Gamma_B^{-1} W_B \Gamma_B^{-1} +e^{2i\xi} \Gamma_B^{-1} W_B \Gamma_B^{-1} W_B \Gamma_B^{-1} + \ldots.
\end{aligned}
\end{equation}
Denote
\begin{equation}
    \psi := W_B\Gamma_B^{-1},
\end{equation}
which is a matrix of dimension $|B|$ with entries $\psi_{zz'} = W_{zz'}/\Gamma_{z'}$ for $z\neq z'$, and zero otherwise. 
The $n$-th coefficient in the expansion is then $\Gamma_B^{-1} \psi^ne^{ni\xi}$. 
Plugging the expansion back in Eq.~\eqref{eq:gen-function-activity} yields 
\begin{equation}\label{39817981798171}
G(\xi) =  C_{x_A \to y_A} \sum_{n=0}^\infty  
    \langle y_A| W_{AB} \Gamma_B^{-1} \psi^n W_{BA} |x_A\rangle  e^{i (n+2) \xi}.
\end{equation}
The support of the distribution is now readily identified to be $n+2$, where $n = 0,1,2,\ldots$. 
This is by construction because an excursion must consist of at least two jumps, corresponding to $A\to B \to A$. 
The coefficients multiplying $e^{i(n+2)\xi}$ give us the corresponding probabilities:
\begin{equation}\label{P_activity}
    P\big(\hat{\mathcal{A}} = n+2\big) = C_{x_A \to y_A}~\langle y_A| W_{AB} \Gamma_B^{-1} \psi^n W_{BA} |x_A\rangle.
\end{equation}
One may verify that this is normalized because the matrix $\psi$ satisfies 
$(1-\psi)^{-1} = \Gamma_B M$. 
Hence, summing Eq.~\eqref{P_activity} over all $n$ yields a factor that precisely cancels the normalization factor $C$.

The average number of transitions in an excursion is 
\begin{equation}
\begin{aligned}
    E\big(\hat{\mathcal{A}}\big) &= \sum_{n=0}^\infty (n+2) P\big(\hat{\mathcal{A}} = n+2\big)
    \\[0.2cm]
    &= 2 + C_{x_A \to y_A} \langle y_A| W_{AB} \Gamma_{B}^{-1}(1-\psi)^{-2}\psi W_{BA} |x_A\rangle.
\end{aligned}
\end{equation}
We can simplify this further by using $(1-\psi)^{-1} \psi = W_B M$. We then get 
\begin{equation}\label{activity_average}
    E\big(\hat{\mathcal{A}}\big) = 2 + E(\hat{\mathcal{A}}_B),
\end{equation}
where $E(\hat{\mathcal{A}}_B) := C_{x_A \to y_A} \langle y_A| W_{AB} M W_{\!B} M  W_{BA}|x_A\rangle$
is the average number of transitions within region $B$ only.
A similar calculation yields the variance of the number of jumps 
\begin{align}\label{activity_variance}
    {\rm var}\big(\hat{\mathcal{A}}\big) =&\ 
    2C_{x_A \to y_A} \langle y_A| W_{AB} M W_B MW_{\!B}M  W_{BA}|x_A\rangle \notag\\
    &+ E(\hat{\mathcal{A}}_B)\Big(1-E(\hat{\mathcal{A}}_B)\Big). 
\end{align}This result can also be interpreted as an application of the law of total variance.

\subsection{Counting a set of specific transitions}
\label{subsec:specific}

Another common application of counting variables in excursions is to count how many times some specific set of transitions took place. 
In particular, consider a set of $r$ transitions between arbitrary pairs of states $z_j'\to z_j$ in $B$ (with $j =1,2,\ldots, r$). 
We define $r$ counting observables $\hat{Q}_j$ as per Eq.~\eqref{counting_observable}, with $\nu_{z_j z_j'} = 1$ and all other $\nu_{xy} = 0$. 
Then, $\hat{Q}_j = n_j$ means that the transition $z_j'\to z_j$ occurred $n_j$ times during an excursion.  
Because we are only counting events within $B$, the block off-diagonals transition matrices~\eqref{tilted_transition_matrix} are untilted
$W_{AB\bm{\xi}}=W_{AB}$, $W_{BA\bm{\xi}}=W_{BA}$. 
Conversely, the $B$ component is tilted as 
\begin{equation}\label{89719817891798711}
    \mathbb{W}_{\!B\bm{\xi}} = \mathbb{W}_{\!B} + \sum_{j=1}^r W_{z_j z_j'} (e^{i\xi_j}-1) |z_j\rangle\langle z_j'|,
\end{equation}
with a distinct counting field $\xi_j$ for each counting observable.

We list the main results here, and discuss how to apply them, but defer some of the technical details to Appendix~\ref{app:monitoring_multiple_transitions}.
Suppose, first, that we are only interested in a single transition $z'\to z$ happening within $B$. 
The corresponding characteristic function can be expanded as
\begin{equation}
    G(\xi) = 1 - \frac{\beta}{1+\gamma} + \sum_{n=1}^\infty \beta  \frac{\gamma^{n-1}}{(1+\gamma)^{n+1}} e^{i n\xi},
\end{equation}
where 
\begin{equation}\label{beta_gamma_single}
\begin{aligned}
    \beta &= C_{x_A \to y_A} W_{zz'} \langle y_A| W_{AB} M |z\rangle\langle z'| M W_{BA}|x_A\rangle, 
    \\[0.2cm]
    \gamma &= W_{zz'} \langle z'|M|z\rangle. 
\end{aligned}
\end{equation}
The support is all non-negative integers, and the corresponding probabilities read 
\begin{equation}\label{single_transition_probs}
\begin{aligned}
    P(\hat{Q} = 0) &= 1 - \frac{\beta}{1+\gamma},
    \\[0.2cm]
    P(\hat{Q} = n) &= \beta \frac{\gamma^{n-1}}{(1+\gamma)^{n+1}},\qquad n = 1,2,3,\ldots.
\end{aligned}
\end{equation}
These therefore provide a very practical and easy to apply method for counting specific transitions. 
All it requires is computing $M$.

Next we discuss the case of monitoring multiple transitions. 
The characteristic function will now have the form (See Appendix~\ref{app:monitoring_multiple_transitions} for a proof):
\begin{align}\label{CF_counting_multiple}
    G(\bm{\xi}) &= 1 - \tr\big\{ \beta (1+\gamma)^{-1}\big\} + \sum_{i=1}^r a_{ii} e^{i \xi_i} + \sum_{i,j=1}^r a_{ji} b_{ij} e^{i(\xi_i + \xi_j)}\nonumber
    \\[0.2cm]
    &\qquad + \sum_{i,j,k=1}^r a_{ki} b_{ij} b_{jk} e^{i(\xi_i + \xi_j+\xi_k)}+\ldots,
\end{align}
where the coefficients are given by
\begin{align}\label{eq:coeffs_ab}
    a_{ji} &= C_{x_A \to y_A} \sqrt{W_{z_i z_i'} W_{z_j z_j'}} \langle y_A|W_{AB} \tilde{M} |z_i\rangle\langle z_j'|\tilde{M} W_{BA}|x_A\rangle, \\[0.2cm]
    b_{ij} &= \sqrt{W_{z_i z_i'} W_{z_j z_j'}} \langle z_i'|\tilde{M}|z_j\rangle. 
\end{align}
Notice that these are all of dimension $r$. The rank of the matrices will be the number of transitions that are being monitored. This holds even if the state space itself is of much higher dimension. 
$\tilde{M}$ is defined as
\begin{equation}
\label{m_tilde}
      \tilde{M} = \left(-\mathbb{W}_{\!B} + \sum_{j=1}^r W_{z_j z_j'} |z_j\rangle\langle z_j'|\right)^{-1}.
\end{equation}
The $n$-th term of the expansion in Eq.~\eqref{CF_counting_multiple} will contain all probabilities with $\sum_j \hat{Q}_j = n$; 
and Eq.~\eqref{single_transition_probs} is recovered if $r=1$. 

To extract the actual probabilities from Eq.~\eqref{CF_counting_multiple} we introduce a set of $r$ projectors $P_j = |j\rangle\langle j|$ of dimension $r$. 
Then
\begin{equation}\label{eq:P_ab}
    P(\hat{Q}_1,\ldots,\hat{Q}_r) = 
    \sum_{\rm all~perms} \tr\Big\{P_1 a (P_1 b)^{n_1} (P_2 b)^{n_2} \ldots (P_r b)^{n_r}\Big\},
\end{equation}
where the sum is over all permutations of how the projectors are inserted in between the matrices $a$ and $b$, which are defined as
\begin{equation}
\label{eq:def-ab}
    a = (1+\gamma)^{-1} \beta (1+\gamma)^{-1}, \qquad 
    b = \gamma(1+\gamma)^{-1}.
\end{equation}
The argument in the left-hand side of Eq.~\eqref{eq:P_ab} is $\hat{Q}_1 = n_1,\ldots,\hat{Q}_r = n_r$.
From these results, we extract meaningful quantities to characterize an excursion.
First, the probability that none of the $r$ transitions happen is 
\begin{equation}\label{multi_all_zero}
    P(\hat{Q}_1=0,\ldots,\hat{Q}_r=0) = 1- \tr\big\{ \beta (1+\gamma)^{-1}\big\}.
\end{equation}
Second, the probability that there are a total of $n$ clicks on all monitored transitions combined, is 
\begin{equation}\label{multi_sum_n}
    P\Big(\sum_{j=1}^r\hat{Q}_j = n\Big) = \tr(a b^{n-1}) = \tr\left\{\beta \gamma^{n-1}(1+\gamma)^{n+1} \right\}.
\end{equation}
Third, the probability that transition $z_1'\to z_1$ happens exactly $n$ times \emph{and} none of the other transitions happen is 
\begin{equation}
    \label{multi_one_n_others_zero}
    P(\hat{Q}_1=n, \hat{Q}_{r\neq 1} = 0) = a_{11} b_{11}^{n-1}.
\end{equation}
From this it then follows that the probability that transition $z_1'\to z_1$ happens (irrespective of how many times it does), but none of the others do, is 
\begin{equation}\label{this_happened_that_didnt}
    P(\hat{Q}_1\neq 0, \hat{Q}_{r\neq 1}=0) = \sum_{n=1}^\infty a_{11} b_{11}^{n-1} = \frac{a_{11}}{1-b_{11}}.
\end{equation}
For example, if $r=2$, this provides a compact answer to the probability that during an excursion some specific transition takes place and another given one transition does not.

\section{Fluctuation theorem and thermodynamic uncertainty relation}\label{sec:FT}

The stochastic entropy production during an excursion was defined in Eq.~\eqref{entropy_production_counting_observable}.
The probability of observing an entropy production $\sigma$  {and the realization $q$ of an arbitrary anti-symmetric counting observable (such as a current)} during an excursion starting from $x_A$ and ending in $y_A$ is given by Eq.~\eqref{Pq_xy}: {
\begin{equation}
\begin{aligned}
    P_{x_A\to y_A}(q,\sigma) = C_{x_A\to y_A}\int \frac{d\xi _qd\xi_\sigma}{(2\pi)^2}
    &\langle y_A| W_{AB \bm{\xi} } M_{\bm{\xi}} W_{BA\bm{\xi}} |x_A\rangle\\[0.2cm]
    &\times~e^{i(q\xi_q+ \sigma\xi_\sigma)},
\end{aligned}
\end{equation}
where $\bm{\xi} = \{\xi_q, \xi_\sigma \}$ and
\begin{equation}
     C_{x_A\to y_A}^{-1} = \langle y_A| W_{AB} M W_{BA} |x_A\rangle.
\end{equation}
Due to the local detailed balance condition~\eqref{detailed_balance}, and the symmetry imposed by time-antisymmetric weights of both $q$ and $\sigma$, the change of variable $\bm{\xi} \to \bm{\xi}' \equiv  \{ -\xi_q, -\xi_\sigma - i \}$ implies that the corresponding tilted matrix in Eq.~\eqref{tilted_transition_matrix} satisfies the special symmetry 
\begin{equation}
    \mathbb{W}_{\bm{\xi}}^{\rm T} = \mathbb{W}_{\bm{\xi}'}.
\end{equation}
Therefore, the probability with the new counting fields reads
\begin{equation}
\begin{aligned}
    P_{x_A\to y_A}(q,\sigma) &= C_{x_A\to y_A}e^{\sigma} \int \frac{d\xi_q d\xi_\sigma}{(2\pi)^2}\\[0.2cm]
    &\times~\langle x_A| W_{AB \bm{\xi} } M_{\bm{\xi}} W_{BA\bm{\xi}} |y_A\rangle e^{-i ( q\xi_q +\sigma \xi_\sigma)},
\end{aligned}
\end{equation}
where we have taken the transpose of the terms in bra-kets.
The integral above is proportional to $P_{y_A\to x_A}(-q,-\sigma)$, except for the normalization constant, hence the fluctuation theorem
\begin{equation}
\label{ft-trajectory-exc}
    \frac{P_{x_A\to y_A}(q,\sigma)}{P_{y_A\to x_A}(-q,-\sigma)} = e^{\sigma} \frac{C_{x_A\to y_A}}{C_{y_A\to x_A}}.
\end{equation}
}The coefficient $C_{x_A\to y_A}/C_{y_A\to x_A}$ is the ratio between the probability of having an excursion $x_A \to y_A$ and having its reverse $y_A \to x_A$. This appears in Eq.~\eqref{ft-trajectory-exc} because the left-hand side denotes a conditional probability on both ends. If region $A$ has a single state, than the ratio $C_{x_A \to y_A}/C_{y_A \to x_A}$ becomes the one and we recover the typical ``exchange'' type fluctuation theorem. 
 {We numerically illustrate our result in Eq.~\eqref{ft-trajectory-exc} with simulated data from excursions in Fig.~\ref{fig:ft-tur}(a).
We emphasize that this result holds for the realization $q$ of any counting observable as long as its weights are anti-symmetric (or equivalently the observable is anti-symmetric under time reversal).

\begin{figure}[t!]
    \centering
    \includegraphics[width=.99\linewidth]{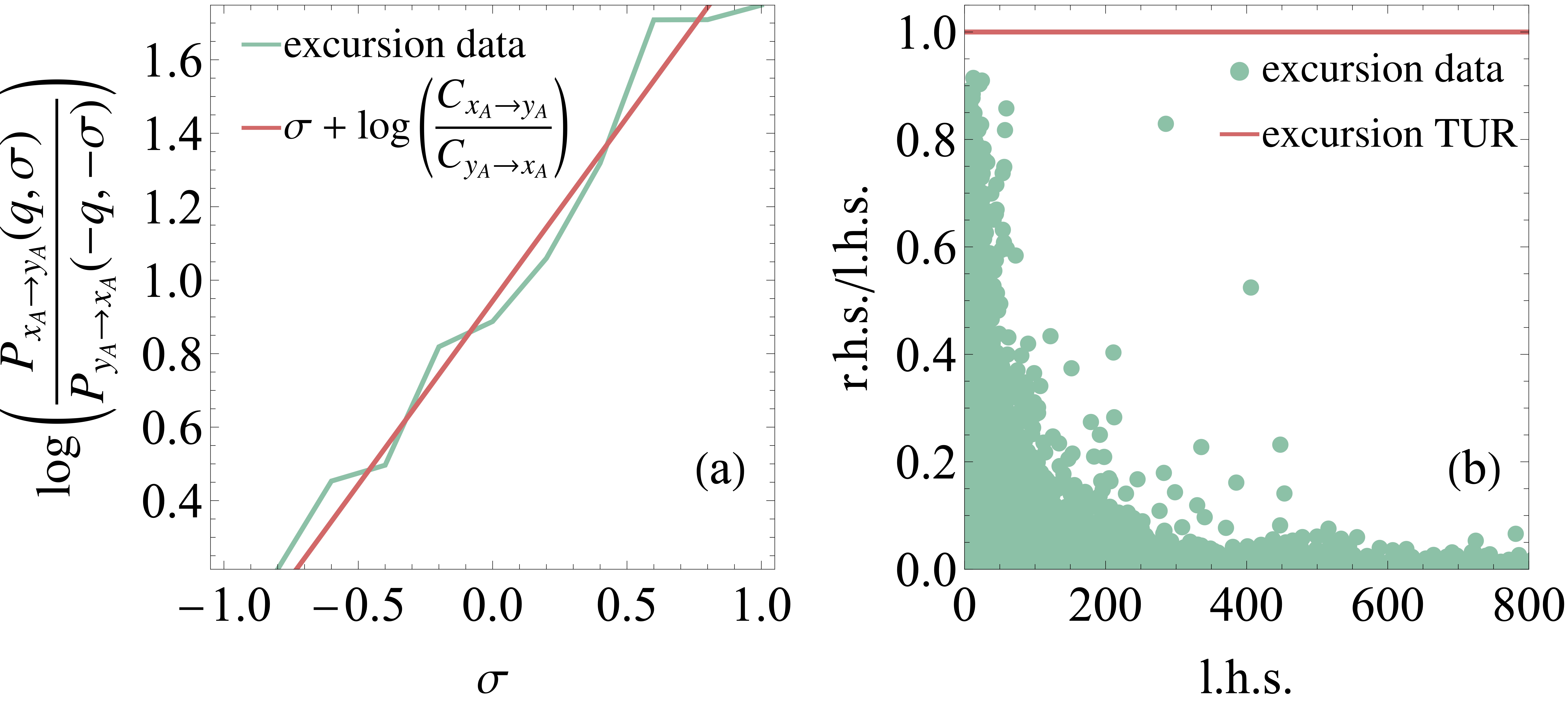}
    \caption{ {Numerical illustration of the (a) excursion fluctuation theorem and (b) the thermodynamic uncertainty relation (TUR). For both plots we considered uniformly randomly generated stochastic transition matrices with entries $W_{ij} \in [0,2], ~ \forall ~i \neq j$ and diagonal elements chosen such that $\sum_i W_{ij} = 0$ to ensure normalization. Region $A$ was taken to have three states and region $B$ to also have three states. For (a) we fixed states $x_A = 1$ and $y_A =3$ to be the initial and final states of the excursions. For (b) we consider an anti-symmetric counting observable randomly generated with entries $\nu_{ij} \in [0,2], ~\forall~ i > j$, with the constraint $\nu_{ij} = -\nu_{ji}$; and 
    l.h.s and r.h.s represent respectively the left hand-side and right hand-side of Eq.~\eqref{eq:TUR}.
    }
    }
    \label{fig:ft-tur}
\end{figure}

Furthermore, it is known that a fluctuation theorem immediately implies a thermodynamic uncertainty relation~\cite{Potts2019, Hasegawa2019b}. In the case of our result in Eq.~\eqref{ft-trajectory-exc}, we obtain
\begin{equation}\label{eq:TUR}
    \dfrac{{\rm var}(\hat{Q})_{x_A \to y_A} + {\rm var}(\hat{Q})_{y_A \to x_A}}{\big(E(\hat{Q})_{x_A \to y_A} +E(\hat{Q})_{y_A \to x_A} \big)^2} \geq \dfrac{1}{e^{[E(\hat{\Sigma})_{x_A \to y_A} +E(\hat{\Sigma})_{y_A \to x_A}]/2}-1},
\end{equation}where averages and variances are conditioned over the initial and final excursion states. A proof of this result is provided in Appendix~\ref{app:TUR}, and numerical verification for the bound over randomly generated transition matrices is provided in Fig.~\ref{fig:ft-tur}(b).
This uncertainty relation is valid at the excursion level, and it recovers the result of Ref.~\cite{Hasegawa2019b} when region $A$ has a single state $x_A = y_A$.
A new insight follows directly from the excursion level fluctuation theorem. It reveals how fluctuations of a current along excursions $x_A \to y_A$ and their reverse $y_A \to x_A$ are bounded by entropy production. 
As with all thermodynamic uncertainty relations, it pinpoints a trade-off between precision and fluctuations. 
However, the result for excursions signifies in a concrete way the interplay between variance and averages of the forward and backward processes, which is not necessarily clear for the generic trajectory-level result.

\section{Applications}
\label{sec:applications}

To illustrate our framework, we apply it to characterize several examples of interest and to showcase the versatility of the methods developed herein. 

\subsection{Simplified absorption refrigerator}
\label{sec:simplified-abs-refrigerator}

Quantum absorption refrigerators have been described in the literature for many years~\cite{Mitchison2019, Levy2012, Levy2012a, Correa2014, Correa2013, Cangemi2024}. 
Recently, they have received a surge of interest due to its experimental implementation~\cite{aamirThermallyDrivenQuantum2025, blokQuantumThermodynamicsQuantum2025, Campbell2025}.
The system consists of three qubits, each connected to their respective reservoirs which are assumed to be bosonic (at different temperatures), denoted $c$ (cold), $h$ (hot), and $w$ (work). 
The desired task is to cool down the cold qubit, so the temperatures are taken to obey $T_c < T_h < T_w$.
Each qubit has a different energy gap given by $\omega_\alpha$ where $\alpha = c,~h,~w$.
The three qubits interact resonantly (i.e. when $\omega_c + ~\omega_w = ~\omega_h$) through an effective three-body interaction, allowing the conversion of one excitation in the work and cold reservoirs into one in the hot reservoir, effectively cooling the cold reservoir in the process, see Fig.~\ref{fig:absorption-refrigerator}(a). 
The behavior of this system at the ensemble level is characterized by the \emph{cooling window}, which provides bounds to the temperatures of the reservoirs such that the system works as a refrigerator:
\begin{equation}
\label{eq:cooling-window}
  \dfrac{\omega_c}{\omega_w} < 
   \dfrac{T_c(T_w - T_h)}{T_w (T_h - T_c)}.
\end{equation}
Note that, since the reservoirs are bosonic, the cooling window~\eqref{eq:cooling-window} can be recast as a function of occupation numbers.

\begin{figure}[t!]
    \centering
    \includegraphics[width=.95\linewidth]{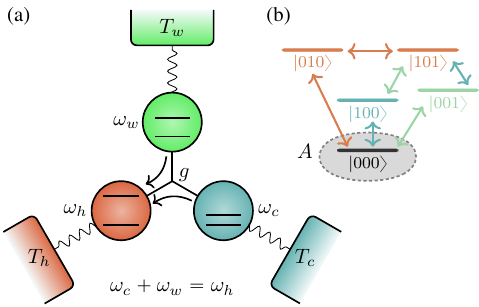}
    \caption{(a) Diagram of the three-qubit absorption refrigerator, each qubit is connected to its respective heat reservoir and has a different energy gap. Labels are $c$ for cold, $h$ for hot, and $w$ for work. The resonant interaction $g$ couples the states $|010\rangle \leftrightarrow |101\rangle$ and takes place only if $\omega_c + \omega_w = \omega_h$.
    (b) Energy landscape of the simplified three-qubit refrigerator model, where the arrows indicate the possible transitions. Colors indicate which reservoir induced the transition. Region $A$ consists of the ground state (gray) and $B$ of all excited states (colored). We adopt the convention ``$chw$'' to denote the states.}
    \label{fig:absorption-refrigerator}
\end{figure}

First, we qualitatively discuss a simplified version of the absorption refrigerator, where only five energy states are considered (in constrast with all eight states). 
In Ref.~\cite{Fiusa2025} we discussed in detail results concerning three counting observables of interest: cold current, entropy production, and dynamical activity.
In the next subsection, we promote a similar discussion for the full model.
We denote the states by $0$ (resp. $1$) representing that the qubit is in the ground state (resp. is excited), and we follow the convention $chw$. Region $A$ is taken to be spanned only by the state $|000\rangle$, where all qubits are in the ground state. 
$B$ is spanned by the other four states: $|001\rangle, ~|010\rangle, ~|100\rangle, ~|101\rangle$ [Fig.~\ref{fig:absorption-refrigerator}(b)].
 {This is the order of states we adopt for our convention when denoting elements of the stochastic transition matrix. For example, the matrix element $W_{14}$ represents the transition rate $|100 \rangle \to |000\rangle$.}
In the parameter regime of Ref.~\cite{aamirThermallyDrivenQuantum2025}, the classical master equation~\eqref{M} approximates the quantum dynamics very well provided that the coupling $g$ is replaced by some classical $g'$, see Appendix~\ref{app:refrigerator}.

The transition among states come from two sources, the reservoirs whose transitions rates are described by $\Gamma_\alpha n_\alpha$ for injections and $\Gamma_\alpha (n_\alpha +1)$ for extractions, and the three body interaction $g'$ that couples the states $|010\rangle \leftrightarrow |101\rangle$ provided that $\omega_h = \omega_c + \omega_w$. 
Since the transitions are induced by a single reservoir, we omit the index $\ell$ in our discussion.
$n_\alpha = \left[\exp({\omega_\alpha}/T_\alpha) - 1\right]^{-1}$ denotes the occupation number and $\Gamma_\alpha$ is the coupling strength between the qubit and the reservoir.
The stochastic transition matrix (without the diagonal elements) is then given by
\[
W =
\begin{pmatrix}
0 & \Gamma_w (n_w + 1) & \Gamma_h (n_h + 1) & \Gamma_c (n_c + 1) & 0 \\
\Gamma_w n_w & 0 & 0 & 0 & \Gamma_c (n_c + 1) \\
\Gamma_h n_h & 0 & 0 & 0 & g' \\
\Gamma_c n_c & 0 & 0 & 0 & \Gamma_w (n_w + 1) \\
0 & \Gamma_c n_c & g' & \Gamma_w n_w & 0
\end{pmatrix}
,\]
where the classical effective three-body interaction strength is 
\begin{equation}
\label{g-cl}
        g'=\frac{4g^2}{\Gamma_c(n_c+1)+\Gamma_h(n_h+1)+\Gamma_w(n_w+1)},
\end{equation}
which follows from perturbation theory~\cite{Prech2023}.

Given the topology of the states and transitions in $B$, this model inherits a nice feature when it comes to constructing the counting observables from Eq.~\eqref{counting_observable}. Namely, the thermodynamical currents such as cold, hot and work currents can only have three possible values per excursion. To see this, consider the cold current that is constructed by setting the weights $\nu_{ xy} = \pm 1$ in Eq.~\eqref{counting_observable} if the transition $x \rightarrow y$ removes (resp. adds) an excitation from (resp. into) the cold reservoir. 
The three possibilities for the cold current are: a \emph{success} where cooling took place and the cold current is $\hat{J}_c = 1$, e.g. $|000\rangle \to |100\rangle \to |101\rangle \to |010\rangle \to |000\rangle$; a \emph{disaster} where heating took place the cold current is $\hat{J}_c = -1$, e.g. $|000\rangle \to |010\rangle \to |101\rangle \to |001\rangle \to |000\rangle$, and a \emph{bounce}/\emph{fail} where $\hat{J}_c = 0$, e.g. $|000\rangle \to |100\rangle \to |000\rangle$. 
This feature also arises for the entropy production, where the positive and negative contributions are related via fluctuation theorems~\eqref{ft-trajectory-exc}.

\subsection{Full absorption refrigerator}
\label{sec:full-abs-refrigerator}
Now we turn our attention to the full model, with all eight states: $|000\rangle, ~|001\rangle, ~|010\rangle, ~|011\rangle, ~|100\rangle, ~|101\rangle, ~|110\rangle, ~|111\rangle$.
The three reservoirs are bosonic with occupation number $n_{\alpha}$ and each qubit is coupled to its reservoir with coupling strength $\Gamma_\alpha$, where $\alpha = c, ~h, ~w$. 
All elements of the transition matrix therefore have the form $\Gamma_\alpha n_\alpha$ and $\Gamma_\alpha (n_\alpha +1)$, except for the transition $|101\rangle \leftrightarrow |010\rangle$, which is denoted $g'$.
The stochastic transition matrix (without the diagonal elements) is then given by 
\begin{widetext}\begin{equation}
\label{eq:W-matrix}
W = \begin{pmatrix}
0 & \Gamma_w (n_w+1) & \Gamma_h(n_h+1) & 0 & \Gamma_c (n_c+1) & 0 & 0 \\
\Gamma_w n_w & 0 & 0 & \Gamma_h(n_h+1) & 0 & \Gamma_c(n_c+1) & 0 & 0 \\
\Gamma_h n_h & 0 & 0 & \Gamma_w(n_w+1) & 0 & g' & \Gamma_c(n_c +1) & 0 \\
0 & \Gamma_h n_h  & \Gamma_w n_w & 0 & 0 & 0 & 0 & \Gamma_c (n_c+1) \\
\Gamma_c n_c & 0 & 0 & 0 & 0 & \Gamma_w (n_w+1) & \Gamma_h (n_h+1) & 0 \\
0 & \Gamma_c n_c & g' & 0 & \Gamma_w n_w & 0 & 0 & \Gamma_h (n_h +1) \\
0 & 0 & \Gamma_c n_c & 0 & \Gamma_h n_h & 0 & 0 & \Gamma_w (n_w + 1) \\
0 & 0 & 0 & \Gamma_c n_c & 0 & \Gamma_h n_h & \Gamma_w n_w & 0 \\
\end{pmatrix},\end{equation}
\end{widetext}where the renormalized coupling $g'$ is given by Eq.~\eqref{g-cl}. To characterize the counting observables, we carry out a similar analysis to the one done in Ref.~\cite{Fiusa2025}. 

\begin{figure}
    \centering
    \includegraphics[width=0.99\linewidth]{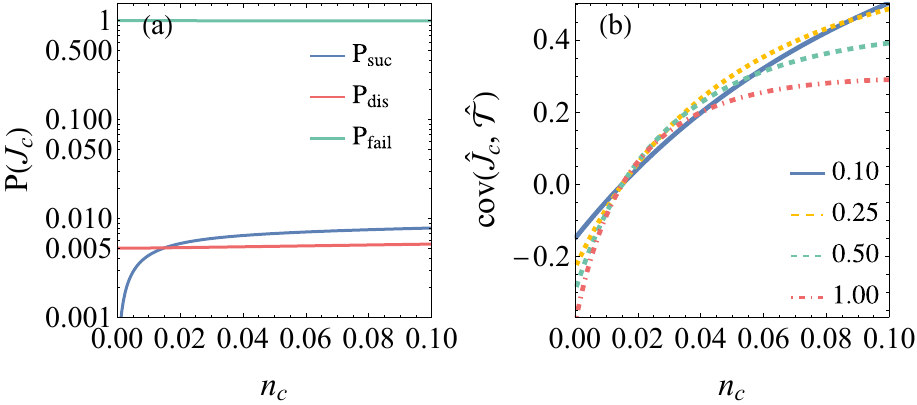}
    \caption{(a) Probabilities for the cold current within an excursion as a function of cold bath occupation number $n_c$. $P_\textrm{suc}$ represents $P(\hat{J}_c = 1)$,  $P_\textrm{dis}$ represents $P(\hat{J}_c = -1)$, and  $P_\textrm{fail}$ represents $P(\hat{J}_c = 0)$.
    (b) Covariance between cold current and excursion duration as a function of $n_c$, for different values of $\Gamma_c$ with fixed $g = 10$.
    For all plots, we fixed $\Gamma_h = 50, \ \Gamma_w = 0.5, \ n_h = 0.005,\ n_w = 0.5$.
    Parameters are well in accordance with the parameter regime of \cite{aamirThermallyDrivenQuantum2025} and the classical approximation.}
    \label{fig:plots-refrigerator}
\end{figure}

First, we begin with the cold current which is constructed by setting the weights $\nu_{xy} = \pm 1$ in Eq.~\eqref{counting_observable} if the transition $x \rightarrow y$ removes (resp. adds) an excitation from (resp. into) the cold reservoir. 
To study the cold current we take region $A$ to be spanned by the states $|000\rangle,~|010\rangle,~|001\rangle,~|011\rangle$.
This is the natural choice for the full refrigerator model, because now the excursions start whenever an excitation enters the cold qubit and end whenever the excitation leaves.
This also assures the nice property that the cold current may only assume three possible values per excursion. 
Indeed, following the counting field expansion [see Eq.~\eqref{Pq_xy}], the probability of the values for the cold current is written as
\begin{equation}
\begin{aligned}
\label{cold-current-expanded}
    P(\hat{J}_c) &= \int_{-\infty}^{\infty} \dfrac{d 
\xi}{2\pi} \left(P_\textrm{suc}~e^{i\xi} + P_\textrm{dis}~e^{-i\xi}+P_\textrm{fail}\right)\\[0.2cm]
&=P_{\rm suc} \delta(\hat{J}_c -1) + P_{\rm fail} \delta(\hat{J}_c-0)+P_{\rm dis} \delta(\hat{J}_c +1),
\end{aligned}
\end{equation}since the span of the cold current is discrete, the integration results in 
the Fourier transform of the Dirac deltas centered around each possible value of $\hat{J}_c$.
Note that the coefficients in the expansion of the counting field are the probabilities that the counting field variable has picked up a particular phase. 
In other words, $P_\textrm{suc}$ is the probability that the cold current has value $\hat{J}_c = 1$ in an excursion. Likewise, $P_\textrm{dis}$ is the probability that the cold current has value $\hat{J}_c = -1$ and $P_\textrm{fail}$ the probability of $\hat{J}_c = 0$. Fig.~\ref{fig:plots-refrigerator}(a) shows the behavior of the probabilities as a function of the cold bath occupation number $n_c$, where it is clear that cooling failures (bounces) dominate the dynamics.
The value where $P_{\rm suc}$ becomes larger than $P_{\rm dis}$ is precisely the value of $n_c$ where the cooling window begins.
In Fig.~\ref{fig:plots-refrigerator}(b), we show the covariance between the cold current $\hat{J}_c$ and the excursion duration $\hat{T}$ as a function of the cold bath occupation $n_c$. Interestingly, longer excursions tend to have larger $\hat{J}_c$ in the cooling window, which is not necessarily the case when $\langle \hat{J}_c \rangle >0$, and once again, a vanishing covariance pinpoints the value of $n_c$ where the cooling window begins.

Equation~\eqref{cold-current-expanded} illustrates another feat of the excursion framework. This is a new paradigm for analyzing the performance of cooling/heating protocols beyond the steady state. 
 {The steady state cold current can be written as function of the probabilites as follows
\begin{equation}
    J_c^{\rm ss} = \dfrac{P_{\rm suc}-P_{\rm dis}}{\mu}.
\end{equation}
This result can also be recovered by computing a generic classical current by solving for the steady state in Eq.~\eqref{M_vec}.
The steady state current takes into account only the difference between the probability of successes and disasters, overlooking how dominant failures are.}
This raises questions for designing useful thermodynamic protocols, e.g. a protocol may have a large \emph{cooling capacity} $P_\textrm{suc} \gg P_\textrm{dis}$ but have low \emph{cooling precision} $P_\textrm{fail} \gg P_\textrm{suc}$.
Non-trivial tradeoffs emerge from this analysis.

For the next two counting observables, we go back to the case where region $A$ is spanned only by the state $|000\rangle$. This allow us to use the results from Eqs.~\eqref{fcs-current} and~\eqref{fcs-diffusion} to characterize fluctuations.
Entropy production $\hat{\Sigma}$ was constructed in Eq.~\eqref{entropy_production_counting_observable}, and we highlighted its importance in our framework when we discussed fluctuation theorems in Sec.~\ref{sec:FT}.
Likewise, the excursion activity $\hat{\mathcal{A}}$ was constructed in Eq.~\eqref{eq:a-activity} and we have provided analytical formulas for its expectation value in Eq.~\eqref{activity_average} and variance in Eq.~\eqref{activity_variance}.
In Figs.~\ref{fig:plots-refrigerator-full}(a),(b) we characterize the diffusion coefficient~\eqref{fcs-diffusion} in terms of its components, both for the entropy production $D_{\Sigma}$ and the dynamical activity $D_{\mathcal{A}}$. In the former, the fluctuations are strongly dominated by the variance term of $\hat{Q}=\hat{\Sigma}$; whereas in the latter there is a clear competition between different contributions, with $E(\hat{\mathcal{A}})^2$ being the most relevant term.
This result sheds light on the very nature of noise in FCS, where the dominant contribution can change depending on the counting observable. $\hat{\Sigma}$ is an observable that can have positive and negative values related by fluctuation theorems, whereas $\hat{\mathcal{A}}$ is strictly positive.
The stark contrast of behavior on the two observables shows the versatility of the excursion framework to characterize fluctuations.

\begin{figure}
    \centering
    \includegraphics[width=0.99\linewidth]{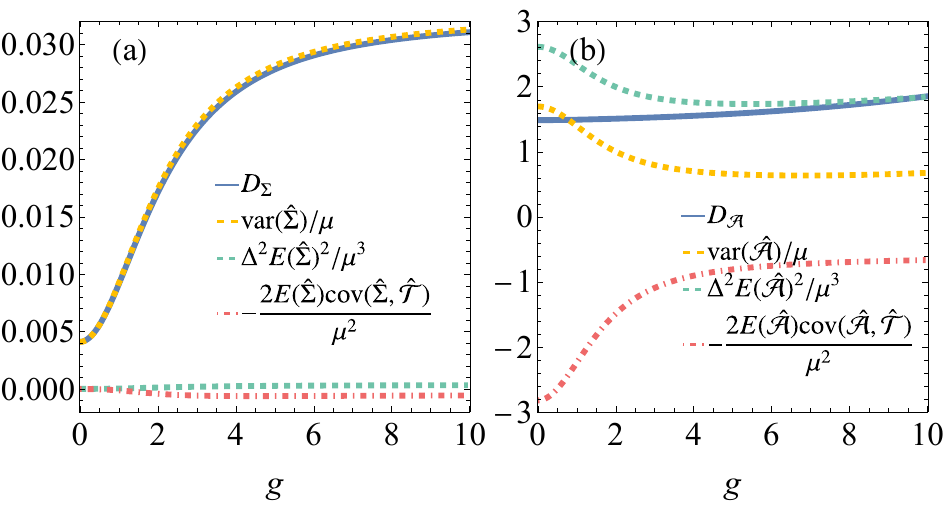}
    \caption{
    Diffusion coefficient (noise) and its decomposition for (a) entropy production, and (b) dynamical activity as a function of $g$, with $n_c = 0.1$ and $\Gamma_c = 0.1$. 
    For all plots, we fixed $\Gamma_h = 50, \ \Gamma_w = 0.5, \ n_h = 0.005,\ n_w = 0.5$.
    Parameters are well in accordance with the parameter regime of \cite{aamirThermallyDrivenQuantum2025} and the classical approximation.}
    \label{fig:plots-refrigerator-full}
\end{figure}

\subsection{Cellular sensing}
\label{sec:cell-sensing}

As another illustration, we use the framework developed herein to explore excursions in models of cellular sensing. Ligands from the environment stochastically bind to membrane-bound receptors of cells, which in turn send downstream signals through a complex biochemical network, and the statistics of these binding events allow for the inference of ligand concentration in the environment. This process is key for cell functions, but it is performed with finite accuracy and incurs energetic costs. Errors in the inferred concentration were shown to be bound from below~\cite{bergPhysicsChemoreception1977, bialekPhysicalLimitsBiochemical2005, Mora2010, Harvey2023}, which involves analyzing the time spent in signaling states, therefore the duration of excursion through such states.

\begin{figure}
    \centering
    \includegraphics[width=0.49\linewidth]{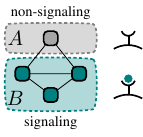}
    \caption{Cellular sensing is often modelled as a stochastic process where some states represent unbound receptors while others represent bound, properties such as the environmental concentration of ligands are learned from the excursions throughout signaling states.}
    \label{fig:sensing-diagram}
\end{figure}

More specifically, in Harvey et al.~\cite{Harvey2023}, the process is modeled as a continuous-time Markov chain partitioned into two subsets---one non-signaling state and multiple signaling ones---a ``simple observer'' infers ligand concentration from the fraction of time spent in signaling states, see Fig.~\ref{fig:sensing-diagram}. A binding event occurs at a rate proportional to the concentration $c$ and causes the system to go from the non-signaling to a signaling state, and the system can jump between distinct signaling states before unbinding. The error of the estimated concentration $\hat{c}$ is shown to satisfy an energy-accuracy trade-off and can be calculated as
\begin{equation}
    \epsilon_{\hat{c}}^2 = \frac{ \mathrm{var}( \hat{c} ) }{ E(\hat{c})^2 } = \frac{2}{\overline{N}} \frac{T_\text{unbind}}{T_\text{hold}},
\end{equation}
where $\overline{N}$ is the mean number of binding events during the observation, $T_\text{unbind}$ is the mean first-passage time from a signaling state to the non-signaling, and $T_\text{hold}$ is the mean time before unbinding given a binding event. Immediately, we observe the important role of excursions for the accuracy in cell sensing, since $T_\text{hold} = E( \hat{ T } )$ is the mean duration of excursions in the sub-set of signaling states, which can be analytically calculated using Eq.~\eqref{average_excursion_time}.

The energy consumed by the cell to sustain the sensing mechanism is defined as the entropy production of the process, which is a counting variable defined over the excursion. Numerous results of the developed stochastic excursions framework immediately apply, including unraveling the diffusion coefficient $D$ in terms of the statistics of $\hat{Q}$ and $\hat{T}$, cf. Eq.~\eqref{fcs-diffusion}.

\begin{figure}
    \centering
\includegraphics[width=0.99\linewidth]{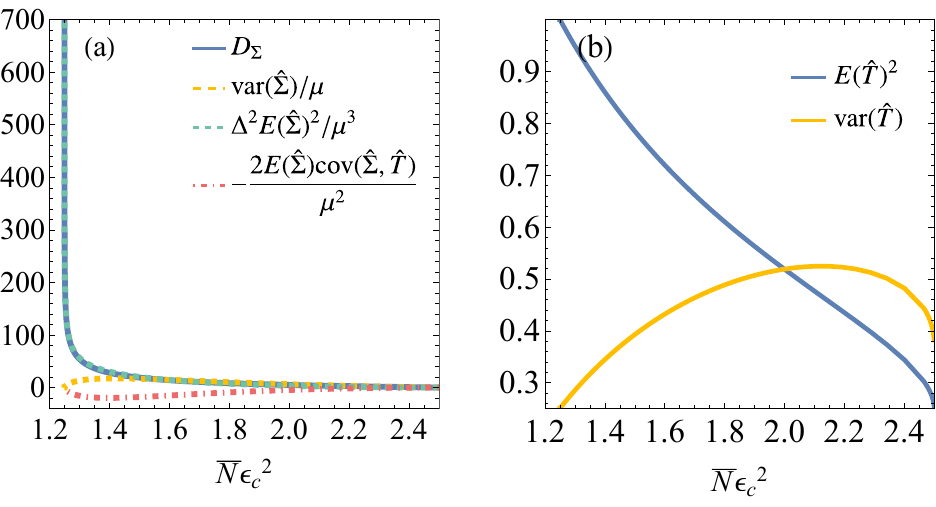}
    \caption{Analysis of excursion properties on a sensing-inspired model. The state-space is a uniform ring with 5 states, one of which being non-signaling, $r_+ = 4$ and $r_-$ varies. 
    (a) Diffusion coefficient (noise) of entropy production and its distinct contributions in terms of the estimation error for $c=1$.
    (b) Moments of excursion duration.}
    \label{fig:sensing}
\end{figure}

We now consider the uniform ring network studied in Ref.~\cite{Harvey2023}, where one state is non-signaling and $n-1$ are signaling. All transitions in one direction have rate $r_+$ while the others have $r_-$. In this model, the energy consumption, measured by entropy production, strongly fluctuates at high-accuracy regimes. Equation~\eqref{fcs-diffusion} allows pinpointing the leading contribution to such large diffusion coefficients. For $n=5$, $r_+ = 4$ and varying values of $r_-$, we observe that $\Delta^2 E(\hat{Q})^2 / \mu^3$ is much larger than the remaining terms when the error is small, as shown in Fig.~\ref{fig:sensing}(a). This behavior, which is robust to values of $c$, is caused by the mean entropy production per excursion $E(\hat{Q})$ being orders of magnitude higher than the contributions that include its variance and the covariance with excursion times.

Another insight enabled by applying the excursion framework concerns the interplay between the mean and variance of the excursion duration, whose relation is particularly important in renewal processes, which is the case when there is a single state in subset $A$. In the same setup as in the previous paragraph, we observe in Fig.~\ref{fig:sensing}(b) that excursion durations have a sub-Poissonian distribution in high-accuracy regimes where $\text{var} (\hat{T}) < E (\hat{T})^2$, while they become super-Poissonian as estimation errors increase, with $\text{var} (\hat{T}) > E (\hat{T})^2$. This is the case for any value of concentration since the moments of $\hat{T}$ do not depend on $c$. A super-Poissonian distribution implies more variability in the number of binding events during an observation window, and slower convergences of empirical quantities.

The robustness of these properties with respect to different topologies and parameters has yet to be exhaustively investigated. Nevertheless, here we aim to showcase the potential of the method to reveal relevant behaviors in systems where excursions play an important role, such as cellular sensing. For instance, the super-Poissonian regime likely has a key contribution to the lower accuracy of concentration estimates since the time spent in signaling states strongly fluctuates, which would be worth investigating in more detail.

\subsection{Birth-and-death processes}
\label{sec:birth-and-death}

To wrap up our examples, we now provide an illustration of some of the analytical calculations that can be carried over with the stochastic excursion framework.
Consider a birth-and-death process defined by a state space $x = 0,1,2,3,\ldots$.
This model nicely illustrates the stochastic excursion framework for a system with infinite space state.
The system can transition upwards with rate $w_{\rm up}$ and downwards with rate $w_{\rm dw}$. 
The transition matrix is thus
\begin{equation}\label{birth_death_W}
    W = \begin{pmatrix}
        0   & w_{\rm dw}   & 0     & 0 &  0 & \ldots \\ 
        w_{\rm up} & 0     & w_{\rm dw}   & 0 &  0 &\ldots \\
        0   & w_{\rm up}   & 0     & w_{\rm dw}&  0 &\ldots \\
        0   & 0   & w_{\rm up}     & 0&  w_{\rm dw} &\ldots \\
        0   & 0   & 0     & w_{\rm up}&  0 &\ldots \\
        \vdots& \vdots & \vdots &\vdots & \vdots & \ddots
    \end{pmatrix}.
\end{equation}
Suppose we select the state $x=0$ as region $A$ and states with $x>0$ as region $B$. 
This kind of birth-and-death process is the prototype for busy periods of single-server queueing models~\cite{Lindley1952, kleinrock1974queueing, cohen1969single, gross2011fundamentals, Bhat2008}.
In this case $w_{\rm up}$ is the rate at which customers arrive and $w_{\rm dw}$ is the rate at which customers are served.
An excursion characterizes a single busy period of the queue, during which the server will be constantly attending to the customers. 

The main parameter that characterizes the behavior of the queue is the ratio of transition rates, which we denote by
\begin{equation}
    r = \frac{w_{\rm up}}{w_{\rm dw}}, 
\end{equation}
the process has a steady state only for $r<1$. 
Otherwise, upward jumps occur more frequently and the average population grows unboundedly. 
The steady state distribution of the dynamics with $W$ matrix given by Eq.~\eqref{birth_death_W} is the geometric distribution
\begin{equation}
    p_x^{\rm ss} = (1-r) r^x. \qquad 
\end{equation}
The normalization constant reads 
\begin{equation}
    C^{-1} = w_{\rm up} (1-r). 
\end{equation} {Note that here we are considering the steady state regime, and thus the normalization constant $C$ does not depend on initial/final states.}

To compute the relevant statistics we need to be able to calculate certain properties of $\mathbb{W}_{\!B}$, which is Toeplitz and tridiagonal.  
Adapting the results in~\cite{da_Fonseca2001}, we find for the $M$ matrix
\begin{equation}\label{birth_and_death_WB_inverse}
    M = \frac{1}{w_{\rm dw}(1-r)}
    \begin{cases}
         1-r^i,& i\leqslant j
        \\[0.2cm]
        r^{i-j}-r^i, & i > j.
    \end{cases}
\end{equation}

More broadly, it might also be convenient to access the entire set of eigenvalues and eigenvalues. 
In this case it is convenient to first take region $B$ to have a finite dimension $d_B$, and then set $d_B\to \infty$ in the end. 
The diagonal form of $W_B$ then reads 
\begin{equation}\label{birth_and_death_eigenstuff}
    \begin{aligned}
        W_{\!B} &= \sum_k \epsilon_k |u_k\rangle\langle v_k|,
        \\[0.2cm]
        \epsilon_k &= 
        2 \sqrt{w_{\rm up} w_{\rm dw}} \cos(k),
        \\[0.2cm]
        |u_k\rangle &= \sqrt{\frac{2}{d_B+1}} \sum_{x=1}^{d_B} r^{(x-1)/2}\sin(kx) |x\rangle,
        \\[0.2cm]
        \langle v_k| &=\sqrt{\frac{2}{d_B+1}} \sum_{x=1}^{d_B} \sin(kx)r^{-(x-1)/2}\langle x|,
        \\[0.2cm]
        k &= \frac{\pi}{d_B+1}, \frac{2\pi}{d_B+1}, \ldots, \frac{d_B\pi}{d_B+1}.
    \end{aligned}
\end{equation}
The eigenvectors of $\mathbb{W}_{\!B} = W_B-\Gamma$ are the same, and the eigenvalues simply have to be shifted by $-\Gamma$. 

After laying down all the necessary tools, we first compute the statistics of the activity, Eq.~\eqref{P_activity}:
\begin{equation}\label{98179818971}
    P(\hat{\mathcal{A}} = n+2) = \frac{w_{\rm dw}}{\Gamma^{n+1}} (\psi^n)_{11}.
\end{equation}
Using the decomposition in Eq.~\eqref{birth_and_death_eigenstuff} we find that 
\begin{align}
   &  {(\psi^n)_{11} = \sum_k \epsilon_k^n \langle 1| u_k\rangle\langle v_k| 1\rangle}\\[0.2cm]
    &= \frac{2}{d_B+1}\sum_k (2\sqrt{w_{\rm up} w_{\rm dw}} \cos k)^n \sin^2k.
\end{align}
The sum can be recast as an integral for large $k$ (equivalently $d_B$), which leads to
\begin{equation}
    (\psi^n)_{11} = (2\sqrt{w_{\rm up} w_{\rm dw}})^n\frac{2}{\pi}\int\limits_0^\pi dk~ \sin^2k \cos^nk. 
\end{equation}
The integral vanishes when $n$ is odd, which is expected: because of the structure of the process, the excursion must always have the same number of jumps upward and downward. 
When $n$ is even we get 
\begin{equation}
    (\psi^{2m})_{11} = \frac{(2m)!}{m!(m+1)!} (\sqrt{w_{\rm up} w_{\rm dw}})^{2m}.
\end{equation}
Plugging this back in Eq.~\eqref{98179818971} yields an analytical formula to evaluate the activity distribution:
\begin{equation}
    P\big(\hat{\mathcal{A}}=2+2m\big)  =\frac{(2m)!}{m!(m+1)!} \frac{r^m}{(1+r)^{2m+1}},\quad 
    m = 0,1,2,\ldots.
\end{equation}
With the distribution at hand, evaluating the average following Eq.~\eqref{activity_average} gives 
\begin{equation}
   E(\hat{\mathcal{A}}) = \frac{2}{1-r}, 
\end{equation}
while variance from Eq.~\eqref{activity_variance} yields
\begin{equation}
    {\rm var}(\hat{\mathcal{A}}) = \frac{4r(1+r)}{(1-r)^3},
\end{equation}
and both diverge when $r\to 1$, as expected.

Next, we compute the statistics of the excursion time following Eq.~\eqref{Pt}. 
Using the eigenstructure of Eq.~\eqref{birth_and_death_eigenstuff}, we get 
\begin{equation}
\label{eq:time-stuff-birth-and-death}
    \langle 1_B |\mathbb{W}_B^2 e^{\mathbb{W}_Bt} |p_B^{\rm ss}\rangle = \sum_k e^{\epsilon_k t} \langle 1_B |\mathbb{W}_B^2 |v_k\rangle \langle w_k |p_B^{\rm ss}\rangle.
\end{equation}
In particular, it follows that
\begin{equation}
    \begin{aligned}
        \langle w_k |p_B^{\rm ss} \rangle &= (1-r)\sqrt{\frac{2}{d_B+1}}\sum_{x=1}^d \sin(kx) r^{(x+1)/2} \\[0.2cm]
        &= (1-r)\sqrt{\frac{2}{d_B+1}}\left(\frac{r \sin(k)}{1+r -2 \sqrt{r}\cos(k)}\right) .   
    \end{aligned}
\end{equation}
The second line is only true in the limit $d_B\to \infty$.
Similarly, 
\begin{equation}
    \langle 1_B |\mathbb{W}_B^2 |v_k\rangle = w_{\rm dw}^2 \sqrt{\frac{2}{d_B+1}} \Big\{(1+r) \sin(k) - r^{1/2} \sin(2k)\Big\}.
\end{equation}
Combining the two results leads to 
\begin{equation}
    \langle 1_B |\mathbb{W}_B^2 |v_k\rangle\langle w_k |p_B^{\rm ss} \rangle = \frac{2w_{\rm dw}^2}{d_B+1}r(1-r) \sin^2(k).
\end{equation}
Therefore we can plug those results in Eq.~\eqref{eq:time-stuff-birth-and-death} and obtain
\begin{equation}
\begin{aligned}
    &\langle 1_B|\mathbb{W}_B^2 e^{\mathbb{W}_B t} |p_B^{\rm ss}\rangle = \frac{2w_{\rm dw}^2}{d_B+1}r(1-r)\sum_k e^{\epsilon_k t} \sin^2(k)
    \\[0.2cm]
    &= w_{\rm dw}^2r(1-r) \frac{2}{\pi}e^{-(w_{\rm up}+w_{\rm dw})t}\int\limits_0^\pi e^{2\sqrt{w_{\rm up} w_{\rm dw}} t \cos(k)}\sin^2 k 
    \\[0.2cm]
    &= 2w_{\rm dw}^2 r(1-r)e^{-(w_{\rm up} + w_{\rm dw})t} \frac{I_1(2\sqrt{w_{\rm up} w_{\rm dw}} t)}{2\sqrt{w_{\rm up} w_{\rm dw}} t},
\end{aligned}    
\end{equation}
where $I_n(x)$ is the modified Bessel function of the first kind.
Combining everything we finally get for Eq.~\eqref{Pt}
\begin{equation}
    P\big(\hat{T}=t\big) = 2w_{\rm up} 
 w_{\rm dw} e^{-(1+r)w_{\rm dw} t} 
    \frac{I_1(2\sqrt{w_{\rm up} w_{\rm dw}} t)}{2\sqrt{w_{\rm up} w_{\rm dw}} t}.
\end{equation}
This is a known result in the theory of queues~\cite{Bhat2008}, describing the statistics of the busy period in an M/M/1 queue. 
The mean excursion time, following Eq.~\eqref{average_excursion_time} is 
\begin{equation}
    E(\hat{T})= \frac{1}{w_{\rm dw}-w_{\rm up}},
\end{equation}
and the variance following Eq.~\eqref{variance_excursion_time} is 
\begin{equation}
    {\rm var}(\hat{T}) = \frac{w_{\rm dw} + w_{\rm up}}{(w_{\rm dw}-w_{\rm up})^2}.
\end{equation}
which again match the results for the M/M/1 queue. 
As expected, we can obtain the results in simple queueing models using excursions, where essentially idle and busy periods alternate just like transitions between regions $A$ and $B$. 
It would be interesting to see if the excursion framework can shed light on more complicated and realistic queueing models and queueing networks.
Nevertheless, a simple queueing model is sufficient to illustrate how systems that have an unbounded number of states can also be described using the excursion framework.

\section{Conclusion}
\label{sec:conclusion}

We have introduced a framework for characterizing counting observables in stochastic Markov jump processes out of equilibrium in Ref.~\cite{Fiusa2025}. By focusing on excursions---trajectories from the inactive phase to the active phase and back---we offer a systematic approach to studying the statistical properties of their times and counting observables. This method generalizes existing efforts in sojourn and first passage times, while also enabling the evaluation of meaningful thermodynamic quantities such as dynamical activity, currents, and entropy production, including their fluctuations.
Systems whose dynamics have a recurring set of states are naturally phrased in terms of excursions leaving this set and then coming back.

The goal of this article is to provide the tools to readily apply the stochastic framework to a wide range of problems of interest. With that in mind, we consolidate the framework of stochastic excursions and provide a comprehensive characterization of counting observables, their statistics, and related quantities.
We derived analytical results for general moments, including moments of the excursion time and counting observables, as well as correlations between excursion duration and counting observables. 
All results can be written in a relatively simple form that depends only on elements of the stochastic transition matrix.
Explicit formulas were provided for two cases of particular interest: activity (the total number of transitions during an excursion) and counting specific transitions. Furthermore, we demonstrated that, in the long-time limit, the average currents and fluctuations derived from excursion statistics are consistent with steady state results obtained from full counting statistics. This connection bridges the gap between trajectory-level dynamics and ensemble-averaged properties.
The steady state result can be re-obtained by statistics of counting observables over excursions.
In particular, the decomposition of the diffusion coefficient into three distinct components provides a new perspective into the nature of noise in classical stochastic systems.
Additionally, we discussed a fluctuation theorem of the exchange type, which holds exactly for individual excursions, showcasing that negative entropy events are exponentially suppressed even at the excursion level.
This result was used to derive a thermodynamic uncertainty relation that illustrates the interplay between forward and backward excursion trajectories.

The stochastic excursion framework also provides insight into finite-time behavior which is overlooked in steady state analyses. 
We illustrated this with the example of the absorption refrigerator, where excursions can not only characterize cooling power but also cooling precision by evaluating the probability of failed cooling cycles. 
In the working parameter range, failed cooling cycles strongly dominate. This characterization is fundamental for designing practical cooling protocols.

Looking ahead, there are many exciting questions to address as follow-ups to the work presented here. One promising direction is the study of correlations between excursions, which may play a key role in the dynamics of systems with memory or feedback. For example, in biological systems, correlations between signaling events could influence the accuracy of sensing or the efficiency of energy transduction. Similarly, in technological applications, such as quantum refrigerators and heat engines, correlations between cooling/heating cycles could affect performance and precision. These correlations appear when $\vert A \vert > 1$, and might yield a more intricate connection with FCS quantities; exploring them within the excursion framework could yield new insights into the design and optimization of such systems.

Another important direction is the extension of our formalism to quantum systems. While the dynamics considered here are purely classical, the concept of excursions is general and could be applied to quantum stochastic processes. 
Finally, excursions could provide new tools for addressing open problems in stochastic thermodynamics, such as the derivation of generalized fluctuation theorems and uncertainty relations. These areas are rich with theoretical challenges and practical applications, and we believe that excursions could offer a fresh perspective. For example, by decomposing entropy production into contributions from individual excursions, one could derive tighter bounds on dissipation or refine existing thermodynamic uncertainty relations. Such advances would deepen our understanding of nonequilibrium processes and provide new strategies for controlling and optimizing small-scale systems.

\begin{acknowledgments}
GF acknowledges great discussions with Tan Van Vu, Jann van der Meer, and Jiheng Duan. This research is primarily supported by the U.S. Department of Energy (DOE), Office of Science, Basic Energy Sciences (BES) under Award No. DE-SC0025516. PH was supported by the project INTER/FNRS/20/15074473 funded by F.R.S.-FNRS (Belgium) and FNR (Luxembourg), and by the European Union (ERC-SuperStoc-101117322).
\end{acknowledgments}

\appendix 

\section{Proof of results in the main text}\label{app:proofs}

\subsection{Evaluation of the excursion duration PDF [Eq.~\eqref{Pt}]}
\label{app:proofs-tilted}
Using the $\delta$-function identity $\delta(t) = \int \tfrac{du}{2\pi}e^{i t u}$ we get 
\begin{equation}
\begin{aligned}
    P\big(\hat{T}=t\big) 
    &= \sum_{N=1}^\infty \sum_{z_1,\ldots,z_N} \int\limits_0^\infty d\tau_1 \ldots \int\limits_0^\infty d\tau_N\\[0.2cm]
    &\times P_{\tt exc}(x_A\to z_{123\ldots} \to y_A)\delta\Big(t - \sum_{j=1}^N \tau_j\Big) 
    \\[0.2cm]
    &= \int\limits_{-\infty}^\infty \frac{du}{2\pi}e^{i t u} 
    \sum_{N=1}^\infty \sum_{z_1,\ldots,z_N} \int\limits_0^\infty d\tau_1 \ldots \int\limits_0^\infty d\tau_N\\[0.2cm]
    &\times P_{\tt exc}(x_A\to z_{123\ldots} \to y_A) e^{-i u (\tau_1+\ldots+\tau_N)},
\end{aligned}
\end{equation} {where $P_{\tt exc}$ was defined in Eq.~\eqref{Pexc}. Note that this result is independent of the number of transitions $N$ within one excursion, because we take into account all possible excursion lengths by summing over $N$.
Computing the integrals over time provides:}
\begin{equation}
\begin{aligned}
P(\hat{T} = t) &= C_{x_A \to y_A} \int\limits_{-\infty}^\infty \frac{du}{2\pi}e^{i t u} \sum_{N=1}^\infty 
    \langle y_A|W_{AB}  {(\Gamma_B+iu)^{-1}}\\
    &\times \big[ W_B  {(\Gamma_B + i u)^{-1}}]^{N-1} W_{BA}|x_A\rangle
    \\[0.2cm]
    &= C_{x_A \to y_A} \int\limits_{-\infty}^\infty \frac{du}{2\pi}e^{i t u} 
    \langle y_A|W_{AB} (iu - \mathbb{W}_{\!B})^{-1} W_{BA}|x_A\rangle
    \\[0.2cm]
    &= C_{x_A \to y_A}  ~\langle y_A|W_{AB} e^{\mathbb{W}_{\!B} t} W_{BA}|x_A\rangle,
\end{aligned}
\end{equation}
which is Eq.~\eqref{Pt}.

\subsection{General formula for the joint probability [Eq.~\eqref{Pqt}] }
\label{app:proofs-joint-prob}

For simplicity we assume a single counting observable, the generalization for any number is straightforward.  {Assuming that the excursion starts in some state $x_A \in A$ and ends in another $y_A \in A$}, start from the stochastic trajectory of an excursion in Eq.~\eqref{Pexc}.
The distribution of $\hat{Q}$ and $\hat{T}$ is 
\begin{equation}
\begin{aligned}
     P(q,t) =&\ \sum_{N=1}^{\infty} \sum_{z_1,\ldots,z_N \in B}\int\limits_0^{\infty} d\tau_1\ldots\int\limits_0^\infty d\tau_N\\[0.2cm]
     &\times P_{\tt exc} 
     \delta\Big(q-\nu_{y_A z_N}-\nu_{z_N z_{N-1}} -\ldots - \nu_{z_2 z_1}-\nu_{z_1 x_A}\Big)\\[0.2cm] 
    &\times \delta\Big(t-\sum_j \tau_j\Big).
\end{aligned}
\end{equation}
we now introduce Fourier representations for the two Dirac delta functions which turns this into 
\begin{equation}
\begin{aligned}
    P(q,t) =&\ C_{x_A \to y_A}\int\limits_{-\infty}^\infty \frac{d\xi}{2\pi}
    e^{i \xi q}
    \int\limits_{-\infty}^\infty \frac{du}{2\pi} 
    e^{i u t}\\[0.2cm]
    & \times
    \sum_{N=1}^\infty \sum_{z_1,\ldots,z_N \in B}\int\limits_0^\infty d\tau_1\ldots\int\limits_0^\infty d\tau_N  W_{y_A z_N} e^{-i\nu_{y_A z_N} \xi} 
    e^{-(\Gamma_{z_N}+i u)\tau_N}\\[0.2cm]
    &\times W_{z_N z_{N-1}} e^{-i \nu_{z_N z_{N-1}} \xi} \ldots  e^{-(\Gamma_{z_1}+i u)\tau_1} W_{z_1 x_A} e^{-i \nu_{z_1 x_A}\xi}.
    \end{aligned}
\end{equation}
Carrying out the integrals over $\tau_j$ yields 
\begin{equation}
\begin{aligned}
    P(q,t) =&\ C_{x_A \to y_A}\int\limits_{-\infty}^\infty \frac{d\xi}{2\pi}
    e^{i \xi q}
    \int\limits_{-\infty}^\infty \frac{du}{2\pi} 
    e^{i u t}\\[0.2cm]
    &\times 
    \sum_{N=1}^\infty \sum_{z_1,\ldots,z_N \in B}
     W_{y_A z_N} e^{-i\nu_{y_A z_N} \xi} 
     \frac{1}{\Gamma_{z_N}+i u}\\[0.2cm]
     &\times
     W_{z_N z_{N-1}} e^{-i \nu_{z_N z_{N-1}} \xi} \ldots  
     \frac{1}{\Gamma_{z_1}+i u}
     W_{z_1 x_A} e^{-i \nu_{z_1 x_A}\xi}.
   \end{aligned}  
\end{equation}
In terms of the tilted matrices in Eq.~\eqref{tilted_transition_matrix} we can write this more compactly as 
\begin{equation}
\begin{aligned}
    &P(q,t) =\ C_{x_A \to y_A} \int\limits_{-\infty}^\infty \frac{d\xi}{2\pi}
    e^{i \xi q}
    \int\limits_{-\infty}^\infty \frac{du}{2\pi} 
    e^{i u t}\\[0.2cm]
    &\times 
    \sum_{N=1}^\infty \langle y_A| W_{AB\xi}(\Gamma_B+iu)^{-1} \left[W_{\!B\xi} (\Gamma_B+iu)^{-1}\right]^{N-1} W_{BA\xi} |x_A\rangle.
\end{aligned}
\end{equation}
Carrying out the sum over $N$ we further get 
\begin{equation}
\begin{aligned}
    P(q,t) =&\ C_{x_A \to y_A} \int\limits_{-\infty}^\infty \frac{d\xi}{2\pi}
    e^{i \xi q}
    \int\limits_{-\infty}^\infty \frac{du}{2\pi} 
    e^{i u t}\\[0.2cm]
    &\times
    \langle y_A| W_{AB\xi} \left(iu - \mathbb{W}_{\!B\xi}\right)^{-1}W_{BA\xi} |x_A\rangle,
\end{aligned}
\end{equation}
where we used $\mathbb{W}_{\!B\xi} = W_{\!B\xi} - \Gamma_B$. 
Carrying out the integral over $u$ and noting that the matrix $\mathbb{W}_{\!B\xi}$ has all eigenvalues with negative real part, we find 
\begin{equation}
    P(q,t) = C_{x_A \to y_A} \int\limits_{-\infty}^\infty \frac{d\xi}{2\pi}
    e^{i \xi q} \langle y_A| W_{AB\xi} e^{\mathbb{W}_{\!B \xi} t} W_{BA\xi} |x_A\rangle,
\end{equation}
which is Eq.~\eqref{Pqt}. 
 {Marginalizing this result over the counting fields immediately provides Eq.~\eqref{Pt}.}

\subsection{Formulas for current [Eq.~\eqref{fcs-current}] and noise [Eq.~\eqref{fcs-diffusion}]}
\label{app:proofs-J-D}

As discussed in the main text, for sufficiently large $t$, the FCS counting observable $\hat{\mathcal{Q}}$ related to a specific interval $[0,t]$ is related to the counting observables $\hat{Q}_n$ of different excursions according to 
\begin{equation}
\hat{\mathcal{Q}} \simeq \sum_{n=1}^{\hat{\mathcal{N}}(t)} \hat{Q}_{n},
\end{equation}
where 
$\hat{\mathcal{N}}(t)$ is the number of excursions that took place in the interval $[0,t]$. 
Here it is assumed that region $A$ has only one state $x_A$, which means $\hat{\mathcal{N}}(t)$ is a renewal process and its asymptotic distribution for large $t$ is Gaussian~\cite{cox1967renewal}.
Finally, there is also the global constraint $\sum_{n=1}^{\hat{\mathcal{N}}(t)} \hat{T}_n^{\rm cyc} \simeq  t$, where
$\hat{T}_n^{\rm cyc} = \hat{T}_n + \hat{\tau}_{x,n}$ is the total cycle time of the $n$-th excursion, including the duration $\hat{T}_n$ of the $n$-th excursion and  the random time $\hat{\tau}_{x,n}$ the system spends in $x$ before an excursion starts. 

The distribution of $\hat{\mathcal{Q}}$ is therefore 
\begin{equation}
\label{steadystate-setup}
    P\bigl(\hat{\mathcal{Q}} = q, t\bigr) = \sum_{N} P(\hat{Q}_{1}+\ldots + \hat{Q}_{N} = q,\hat{T}_1^{\text{cyc}} +\ldots + \hat{T}_{N}^{\text{cyc}} = t). 
\end{equation}
Because region $A$ contains only a single state, the variables $(\hat{Q}_n,\hat{T}_n^{\rm cyc})$ are statistically independent for $n\neq m$. 
We may therefore write 
\begin{align}
&P(\hat{Q}_{1}+\ldots + \hat{Q}_{N} = q,\hat{T}_1^{\rm cyc}  + \ldots + \hat{T}_{N}^{\rm cyc} = t)  \notag \\ 
&= \int
d\hat{Q}_{1}\ldots d\hat{Q}_N \ d\hat{T}_1^{\rm cyc} \ldots d\hat{T}_{N}^{\rm cyc} P(\hat{Q}_{1},\hat{T}_1^{\rm cyc})\ldots P(\hat{Q}_{N},\hat{T}_{N}^{\rm cyc}) \notag\\
&\qquad \times \delta \left(q- \sum_{n=1}^{N}\hat{Q}_{n} \right) \delta \left(t- \sum_{n=1}^{N}\hat{T}_{n}^{\rm cyc} \right).
\end{align}
Here $P(\hat{Q}_n,\hat{T}_n^{\rm cyc})$ is not exactly Eq.~\eqref{Pqt}, only because it refers to $\hat{T}_n^{\rm cyc}$ instead of $\hat{T}_n$. 
But since $\hat{\tau}_{x,n}$ is independent of $\hat{T}_n$, the two are still closely related. 
Using a Fourier representation of the delta function, we can write 
\begin{equation}
\label{steadystate-prob-gen-function}
    P\bigl(\hat{\mathcal{Q}} = q, t\bigr) = \sum_N \int  \dfrac{ d\xi d\omega}{(2 \pi)^2} e^{i(\xi q + \omega t)}G(\xi, \omega)^N,
\end{equation}
where $G(\xi,\omega)$ is the joint characteristic function of $(\hat{Q},\hat{T}^{\rm cyc})$:
\begin{equation}
\label{steadystate-gen-function}
    G(\xi,\omega) = \int d\hat{Q} d\hat{T}^{\rm cyc} e^{-i(\xi q + \omega t)}P(\hat{Q},\hat{T}^{\rm cyc}),
\end{equation}
which is identical for all $n = 1,2,\ldots, N$. 
This can also be written as 
\begin{equation}
\label{steadystate-full-integral}
      P\left(\hat{\mathcal{Q}} = q, t\right) = \sum_N \int  \dfrac{ d\xi d\omega}{(2 \pi)^2} 
      e^{i(\xi q + \omega t) + NC(\xi, \omega)},
\end{equation}
where $C(\xi, \omega) = \log{G(\xi,\omega)}$ is the cumulant generating function.
Conversely, from FCS we know that the LHS of~\eqref{steadystate-full-integral} can be written as \cite{landi2024a}
\begin{equation}
\label{steadystate-result-fcs}
     P\left(\hat{\mathcal{Q}} = q, t\right) = \int \dfrac{d \xi}{2 \pi} e^{i \xi q + K(\xi)t}, \quad
     K(\xi) \simeq -i \xi J - \dfrac{\xi^2}{2}D,
\end{equation}
with $J$ and $D$ denoting the current and diffusion coefficients defined in Eq.~\eqref{FCS_J_D}.
By comparing the two results, we can relate $J$ and $D$ to quantities pertaining to a single excursion. 

To do that, we first expand the cumulant generating function up to second order in $\xi$ and $\omega$, 
\begin{equation}
\begin{split}
\label{steadystate-CGS-expansion}
    C(\xi, \omega) 
    \simeq& -i \omega \mu - i \xi E(\hat{Q}) - \dfrac{\omega^2 \Delta^2}{2}\\
    &\ - \dfrac{\xi^2}{2}\text{var}(\hat{Q}) - \omega \xi \ \text{cov}(\hat{Q},\hat{T}).
\end{split}
\end{equation}
All quantities here are represented in terms of moments computable from  Eq.~\eqref{Pqt}.
Because $\hat{\tau}_x$ is statistically independent from $\hat{Q}$ and $\hat{T}$, it follows that 
${\rm cov}(\hat{Q},\hat{T}^{\rm cyc})={\rm cov}(\hat{Q},\hat{T})$. 
Moreover, we also defined 
$\mu = E(\hat{T}_n^{\rm cyc}) = E(\hat{T}) + \Gamma_x^{-1}$ and $  \Delta^2 = {\rm var}(\hat{T}_n^{\rm cyc}) = {\rm var}(\hat{T}) +\Gamma_x^{-2}$ as the mean and variance of $\hat{T}^{\rm cyc}$.
We then insert Eq.~\eqref{steadystate-CGS-expansion} into~\eqref{steadystate-full-integral} and perform saddle point approximations, first for the integral over $\omega$ then for the sum over $N$. 
Finally, expanding the result to order $\xi^2$ (inside the exponential) we obtain 
\begin{align*}
       P\left(\hat{\mathcal{Q}} = q,t\right) \propto \int & \dfrac{d \xi}{2 \pi} \exp{\left[ i \xi q - \dfrac{i t E(\hat{Q}) \xi}{\mu} - \dfrac{E(\hat{Q})^2 t \Delta^2 \xi^2}{2\mu^3} \right.} \notag \\
       &+ \left.\dfrac{E(\hat{Q}) \text{cov}(\hat{Q},\hat{T}) t \xi^2}{\mu^2}  - \dfrac{\text{var}(\hat{Q})t \xi^2}{2\mu}\right].
\end{align*}
Comparing with $K(\xi)$ from  Eq.~\eqref{steadystate-result-fcs} then yields Eqs.~\eqref{fcs-current} and~\eqref{fcs-diffusion} of the main text.

\subsection{Formulas for monitoring multiple individual transitions [Eq.~\eqref{CF_counting_multiple}]}\label{app:monitoring_multiple_transitions}

Consider that we monitor $r$ specific transitions from $z_j'\to z_j$. 
We begin by splitting the tilted matrix~\eqref{89719817891798711} as 
\begin{equation}
    -\mathbb{W}_{\!B\bm{\xi}} := \tilde{M}^{-1} - H,
\end{equation}
where the matrix $\tilde{M}$ is defined in Eq.~\eqref{m_tilde} and $H$ reads
\begin{equation}
    H(\bm{\xi}) = \sum_{j=1}^r W_{z_j z_j'} e^{i\xi_j} |z_j\rangle\langle z_j'|.
\end{equation}
The characteristic function in Eq.~\eqref{CF} then becomes 
\begin{equation}
    G(\bm{\xi}) = C \langle L| (\tilde{M}^{-1} - G)^{-1} |R\rangle, 
\end{equation}
where 
$\langle L| = \langle y_A|W_{AB}$ and $|R\rangle = W_{BA}|x_A\rangle$ for brevity. 
Using the geometric matrix expansion we get 
\begin{equation}
    G(\bm{\xi}) = C \Big\{ \langle L| \tilde{M} |R\rangle + \langle L|\tilde{M} H \tilde{M} |R \rangle + \langle L |\tilde{M} H \tilde{M} H \tilde{M} |R\rangle + \ldots \Big\}.
\end{equation}
Using the definitions in the main text following Eq.~\eqref{eq:coeffs_ab}, it follows
\begin{equation}
\begin{aligned}
    G(\bm{\xi}) =&\ C \langle L | \tilde{M}|R\rangle + \sum_i a_{ii} e^{i\xi_i} + \sum_{i,j}a_{ji}b_{ij} e^{i (\xi_i +\xi_j)} 
    \\[0.2cm]&+ \sum_{i,j,k} a_{ki}b_{ij}b_{jk} e^{i(\xi_i+\xi_j+\xi_k)} + \ldots,
\end{aligned}
\end{equation}
which is already in the form of Eq.~\eqref{CF_counting_multiple}. 
We still want to futher simplify this result with the goal of casting it in terms only of $M$.
We therefore turn to $\tilde{M}$ in Eq.~\eqref{m_tilde} and rewrite it using the well-known Sherman-Morrison formula~\cite{Sherman1949}:
\begin{equation}\label{appSM_sherman_morrison}
    \Big(A + |z\rangle\langle z'|\Big)^{-1} = A^{-1} - \frac{A^{-1} |z\rangle\langle z'| A^{-1}}{1 + \langle z'|A^{-1}|z\rangle}.
\end{equation}
We apply this result to Eq.~\eqref{m_tilde}. Let 
\begin{equation}
    S_{ij} = \frac{1}{W_{z_iz_i'}}\delta_{ij} + \langle z_i'| M|z_j\rangle, 
\end{equation}
it then results in
\begin{equation}
    \tilde{M} = M - \sum_{ij} S_{ij}^{-1} M |z_i\rangle\langle z_j'|M. 
\end{equation}
Plugging this in Eq.~\eqref{eq:coeffs_ab} allows to express $a$ and $b$ [c.f. Eq.~\eqref{eq:def-ab}] in terms of the coefficients $\beta$ and $\gamma$, which concludes the proof. 

 {
\subsection{Excursion thermodynamic uncertainty relation [Eq.~\eqref{eq:TUR}]}\label{app:TUR}

In this proof, we adapt the steps from Ref.~\cite{Potts2019}, which derived an uncertainty relation for processes with measurement and feedback. The starting point is the fluctuation theorem of Eq.~\eqref{ft-trajectory-exc}, which we write down as
\begin{equation}\label{eq:ft-appendix}
    P_{x_A \to y_A}(q,\sigma) = e^{\sigma + \zeta} P_{y_A \to x_A}(-q, -\sigma),
\end{equation}with $\zeta = \log{(C_{x_A \to y_A}/C_{y_A \to x_A})}$. Consider the (normalized) auxiliary distribution
\begin{equation}\label{eq:aux-distribution}
\begin{aligned}
    R(q,\sigma) &:= \dfrac{P_{x_A \to y_A}(q, \sigma) +  P_{y_A \to x_A}(-q, -\sigma)}{2} \\[0.2cm]
    &= \dfrac{1+e^{-\sigma -\zeta}}{2} P_{x_A \to y_A}(q, \sigma),
    \end{aligned}
\end{equation}whose support is $q, ~\sigma \in \mathbb{R}$. We now compute a few convenient quantities. We make use of the fluctuation theorem in Eq.~\eqref{eq:ft-appendix}, the two forms of Eq.~\eqref{eq:aux-distribution}, the exponential form $\tanh{(x/2)}=(1-e^{-x})/(1+e^{-x})$, and the change of variables $\left\{q,\sigma \right\} \to \left\{-q,-\sigma \right\}$. We start with the first two moments of $\hat{Q}$, where $q$ is the realization of the random variable:
\begin{equation}\label{eq:average-appendix}
\begin{aligned}
    &E(\hat{Q})_R = \int dq d\sigma ~q R(q,\sigma)\\[0.2cm]
    &=\int dq d\sigma~ q  \dfrac{P_{x_A \to y_A}(q, \sigma) +  P_{y_A \to x_A}(-q, -\sigma)}{2}\\
    &=\dfrac{E(\hat{Q})_{x_A \to y_A} - E(\hat{Q})_{y_A \to x_A}}{2}.
\end{aligned}
\end{equation}And
\begin{equation}
      E(\hat{Q}^2)_R =\dfrac{E(\hat{Q}^2)_{x_A \to y_A} - E(\hat{Q}^2)_{y_A \to x_A}}{2}.
\end{equation}The variance therefore reads
\begin{equation}\label{eq:variance-appendix}
    \begin{aligned}
        &\textrm{var}(\hat{Q})_R = E(\hat{Q}^2)_R - E(\hat{Q})_R^2\\[0.2cm]
        &= \dfrac{\textrm{var}(\hat{Q})_{x_A \to y_A} + \textrm{var}(\hat{Q})_{y_A \to x_A}}{2} + \left(\dfrac{E(\hat{Q})_{x_A \to y_A} + E(\hat{Q})_{y_A \to x_A}}{2}\right)^2.
    \end{aligned}
\end{equation}
Lastly, we shall use the following identity:
\begin{equation}
    \begin{aligned}
        E&\left(\left(\hat{Q}-\langle\hat{Q}\rangle_R\right) \tanh{\dfrac{ \hat{\Sigma} + \zeta}{2}} \right)_R = \int dq d\sigma\left( q -\langle\hat{Q}\rangle_R\right)\\[0.2cm]
        &\times~\dfrac{1-e^{-\sigma -\zeta}}{2}P_{y_A \to x_A}(-q,-\sigma)\\[0.2cm]
        &=\dfrac{1}{2}\int dqd\sigma \left( q -\langle\hat{Q}\rangle_R\right)P_{x_A \to y_A}(q,\sigma)\\[0.2cm]
        &-\dfrac{1}{2}\int dqd\sigma \left( q -\langle\hat{Q}\rangle_R\right) P_{y_A \to x_A} (-q,-\sigma)\\[0.2cm]
        &=\dfrac{E(\hat{Q})_{x_A \to y_A} + E(\hat{Q})_{y_A \to x_A}}{2}.
    \end{aligned}
\end{equation}
The same calculation applies to the entropy production, where we conclude
\begin{equation}
     E\left(\left(\hat{\Sigma}+\zeta\right) \tanh{\dfrac{ \hat{\Sigma} + \zeta}{2}} \right)_R = \dfrac{E(\hat{\Sigma})_{x_A \to y_A} + E(\hat{\Sigma})_{y_A \to x_A}}{2}.
\end{equation}
With all the required identities, we are ready to proceed. First step is to apply the Cauchy-Schwarz inequality:
\begin{equation}
\begin{aligned}
    \left(\dfrac{E(\hat{Q})_{x_A \to y_A} + E(\hat{Q})_{y_A \to x_A}}{2} \right)^2 &= E\left(\left(\hat{Q}-\langle\hat{Q}\rangle_R\right) \tanh{\dfrac{ \hat{\Sigma} + \zeta}{2}} \right)_R\\[0.2cm]
    &\leq \textrm{var}(\hat{Q})_R E\left(\tanh{\left( \dfrac{ \hat{\Sigma} + \zeta}{2}\right)^2} \right)_R.
\end{aligned}
\end{equation}The hyperbolic tangent is odd and satisfies $k \tanh{x} \leq \tanh(kx)$ for any $k$ and non-negative $x$. Hence we take $x = (\hat{\Sigma} + \zeta)/2$ and $k = \tanh(\hat{\Sigma} + \zeta)/2$. It then follows
\begin{equation}\label{eq:result-CS}
  \tanh{\left( \dfrac{\hat{\Sigma} + \zeta}{2}\right)^2} \leq \tanh\left[ \tanh\left( \dfrac{\hat{\Sigma} + \zeta}{2}\right) \dfrac{\hat{\Sigma} + \zeta}{2} \right].
\end{equation}
Note that this result holds irrespective of the sign of $\sigma + \zeta$. Back to Eq.~\eqref{eq:result-CS}, we now get
\begin{equation}
\begin{aligned}
       \left(\dfrac{E(\hat{Q})_{x_A \to y_A} + E(\hat{Q})_{y_A \to x_A}}{2} \right)^2 &\leq  \textrm{var}(\hat{Q})_R\\[0.2cm]
       &\times~E\left( \tanh \Big[\tanh{\left( \dfrac{\hat{\Sigma} + \zeta}{2}\right)\dfrac{\hat{\Sigma} + \zeta}{2}\Big] } \right)_R.
\end{aligned}
\end{equation}The argument of the outer tanh is non-negative for any $\{\hat{\Sigma},\zeta \}$, where the function is always concave. Thus we can apply Jensen's inequality, providing
\begin{equation}
\begin{aligned}
       \left(\dfrac{E(\hat{Q})_{x_A \to y_A} + E(\hat{Q})_{y_A \to x_A}}{2} \right)^2 &\leq  \textrm{var}(\hat{Q})_R\\[0.2cm]
       &\times~\tanh \left[ E\left(\tanh{\left( \dfrac{\hat{\Sigma} + \zeta}{2}\right)\dfrac{\hat{\Sigma} + \zeta}{2}} \right)_R\right].
\end{aligned}
\end{equation}Finally, recalling Eqs.~\eqref{eq:average-appendix} and \eqref{eq:variance-appendix}, and using the exponential form of the hyperbolic tangent, all that is let is elementary algebraic manipulations. We then arrive at
\begin{equation}
       \dfrac{{\rm var}(\hat{Q})_{x_A \to y_A} + {\rm var}(\hat{Q})_{y_A \to x_A}}{\big(E(\hat{Q})_{x_A \to y_A} +E(\hat{Q})_{y_A \to x_A} \big)^2} \geq \dfrac{1}{e^{[E(\hat{\Sigma})_{x_A \to y_A} +E(\hat{\Sigma})_{y_A \to x_A}]/2}-1},
\end{equation}which is the thermodynamic uncertainty relation at the excursion level.}

\section{Further comments on the absorption refrigerator}\label{app:refrigerator}

The microscopic Hamiltonian of the model is given by
\begin{equation}
    H = \sum_{\alpha=c,h,w}\omega_\alpha\sigma_\alpha^+ \sigma_\alpha^- + g(\sigma_c^+ \sigma_h^- \sigma_w^+ + \sigma_c^- \sigma_h^+ \sigma_w^-).
\end{equation}
The transitions are induced by the three reservoirs and the interaction term with coupling strength $g$. We assume that the reservoirs are bosonic and thus follow a Bose-Einstein distribution, where occupation numbers are given by $n_\alpha = (\exp(\omega_\alpha/T_\alpha)-1)^{-1}$. The argument is the same in the case where the reservoirs are fermionic. The evolution of the system follows, under the typical reasonable assumptions such as weak coupling, the master equation
\begin{equation}
\label{eq:quantum-me}
    \dfrac{\partial \rho}{\partial t} = \mathcal{L}\rho = -i[H,\rho] + \mathcal{D}(\rho),
\end{equation}where the dissipators are given by
\begin{equation}
   \mathcal{D}(\rho) = \sum_{\alpha=c,h,w} \Gamma_\alpha\left[L_\alpha \rho L_\alpha^\dagger - \dfrac{1}{2}\{L_\alpha^\dagger L_\alpha, \rho\} \right],
\end{equation}with jump operators $L_\alpha = n_\alpha \sigma_\alpha^-$ and $L_\alpha^\dagger = (n_\alpha+1)\sigma_\alpha^+$. 

\begin{figure}
    \centering
    \includegraphics[width=0.99\linewidth]{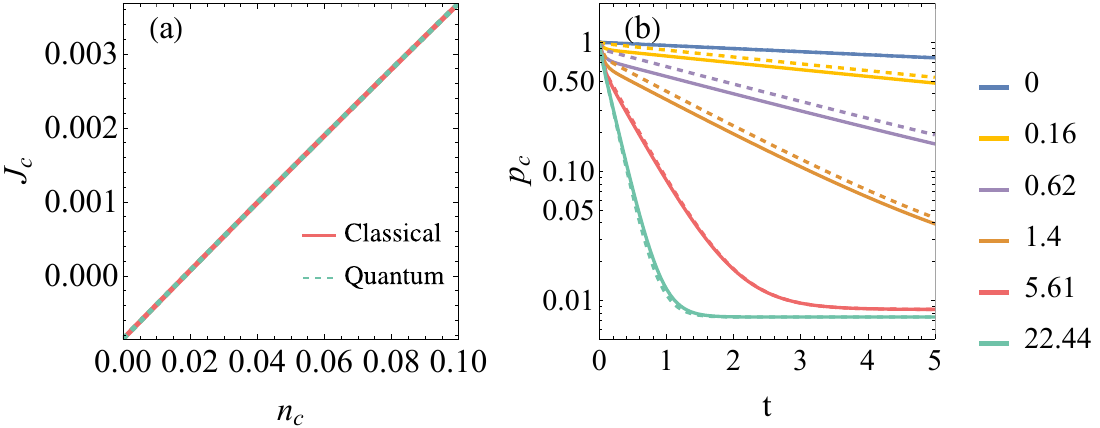}
    \caption{(a) Cold current as a function of cold reservoir occupation number $n_c$, for the classical and quantum dynamics. We considered $n_w = 0.62$ fixed.
    (b) Population of the cold qubit over time for different values of $n_w$, with fixed $n_c = 0.028$.
    Solid lines are obtained from the classical master equation and dashed lines from the quantum master equation.
    All parameters were chosen to match exactly the ones in Ref.~\cite{aamirThermallyDrivenQuantum2025}. We fixed, in units of MHz/$2\pi$, the parameters: $\Gamma_h = 7.2, ~\Gamma_w = 0.07,~\Gamma_c = 0.0087,~ g = 3.2,~\omega_w =  5.327\times 10^3,~\omega_c =  3.725 \times 10^3,~\omega_h = \omega_w + \omega_c$. We fixed the occupation number of the hot reservoir in both plots to be $n_h = 0.007$.}
    \label{fig:classical-quantum-refrigerator}
\end{figure}

From the shape of Eq.~\eqref{eq:quantum-me} and the typical classical master equation~\eqref{M_vec}, it is clear that in certain situations one should be able to map one into the other. 
In the weak coupling and high temperature limit, the jump operators \(L_\alpha\) can be related to transition rates between energy eigenstates. Let us assume the system Hamiltonian \(H\) has eigenstates \(|n\rangle\) with eigenvalues \(E_n\):
\begin{equation*}
 H |n\rangle = E_n |n\rangle.   
\end{equation*}
The jump operators can therefore be expressed in terms of transition rates between these eigenstates. For example, if \(L_\alpha = \sqrt{\gamma_{nm}} |n\rangle \langle m|\), then the Lindblad equation becomes:
\begin{equation}
\frac{d\rho}{dt} = -i[H, \rho] + \sum_{n,m} \gamma_{nm} \left( |n\rangle \langle m| \rho |m\rangle \langle n| - \frac{1}{2} \{|m\rangle \langle m|, \rho\} \right).
\end{equation}
If we assume that the density matrix \(\rho\) is approximately diagonal in the energy eigenbasis, i.e., \(\rho \simeq \sum_n p_n |n\rangle \langle n|\), where \(p_n = \langle n| \rho |n \rangle\) are the populations, the coherences are small and decay rapidly. The Lindblad equation then assumes the form
\begin{equation}
\frac{dp_n}{dt} = \sum_m \left( \gamma_{nm} p_m - \gamma_{mn} p_n \right),
\end{equation}which is precisely (up to notation) the classical master equation in Eq.~\eqref{M}. 

Next, we show how to evaluate the classical and quantum currents at the steady state. This will be used to check the consistency of the classical approximation to describe the absorption refrigerator model. 
The classical current is trivially obtained by setting the proper weights $\nu_{\ell xy}$ at the steady state, which in turn is recovered by setting $dp_x/dt = 0$ in Eq.~\eqref{M}. With the steady state distribution, currents are built by considering
\begin{equation}
\label{eq:classical-current}
    J_{\rm class} = \sum_{\ell x y} \nu_{\ell xy} W_{\ell xy} p_y^{\rm ss}.
\end{equation}Hence, one obtains the cold current by taking weights $\nu_{\ell xy} = \pm 1$ in Eq.~\eqref{counting_observable} if the transition $y \rightarrow_\ell x$ removes (resp. adds) an excitation from (resp. into) the cold reservoir. Those weights are exactly the same ones for the cold current at the excursion level. Likewise, the quantum case follows from a straightforward calculation from Eq.~\eqref{eq:quantum-me}, see e.g.~\cite{Mitchison2019}. One finds
\begin{equation}
\label{eq:quantum-current}
    J_{\rm quant} = 2g \omega_\alpha ~{\rm Im}\left[\tr{(\sigma_c^+ \sigma_h^- \sigma_h^+ \rho^{\rm ss}} )\right], 
\end{equation}where $\rho^{\rm ss}$ is obtained by setting the right-hand side of Eq.~\eqref{eq:quantum-me} to zero ($\mathcal{L}\rho^{\rm ss} = 0$), and $\alpha$ labels the current (the hot current would have a minus sign due to the convention we adopted). The cold current is obtained by choosing $\omega_\alpha = \omega_c$. 

The next step is to verify when the jump operators can be well approximated by the fully decohered version, and when the classical and quantum currents match. Physically, this happens for weak coupling, high temperatures and fast decoherence. 
In practice, we solve the dynamics using the quantum master equation in Eq.~\eqref{eq:quantum-me} and the classical master equation (with $W$ matrix shown in Eq.~\eqref{eq:W-matrix}) in Eq.~\eqref{M} and plot the population of the cold qubit $p_c$ over time in Fig.~\ref{fig:classical-quantum-refrigerator}(b).
We then compare the classical current~\eqref{eq:classical-current} with the quantum~\eqref{eq:quantum-current} in Fig.~\ref{fig:classical-quantum-refrigerator}(a).

Clearly, the results for the current $J_c$ and the real-time dynamics of the population of the cold qubit $p_c$ suggest that indeed the classical master equation~\eqref{M} is a very good approximation of the full quantum dynamics for the three qubit absorption refrigerator. We emphasize that the parameters used in the plots of Fig.~\ref{fig:classical-quantum-refrigerator} were chosen to match the experimental values of Ref.~\cite{aamirThermallyDrivenQuantum2025}. Of course, the parameter regime used in the simplified model of Ref.~\cite{Fiusa2025} and the full model in this paper [see results in Figs.~\ref{fig:plots-refrigerator} and~\ref{fig:plots-refrigerator-full}] also work very well in the classical approximation.

\bibliography{letter}
\bibliographystyle{ieeetr}

\end{document}